\documentclass[12pt]{iopart}   
\usepackage{iopams}
\usepackage{epsf}
\usepackage{euscript}
\usepackage{graphicx}

\renewcommand{\footnoterule}

\begin{document}
\title {Counting nodal domains on surfaces of revolution}
\author{Panos D. Karageorge$^{2,3}$
and Uzy Smilansky$^{1,2}$} \address{$^1$Department of Physics of
Complex Systems, The Weizmann Institute of Science, Rehovot 76100,
Israel.} \address{$^2$School of Mathematics, University of Bristol, Bristol BS8 1TW, UK.}
\address{$^3$Department of Physics, University of Crete, Heraklion 71003, Greece.}

\date{\today}

\begin{abstract}
We consider eigenfunctions of the Laplace-Beltrami operator on
special surfaces of revolution. For this separable system, the nodal
domains of the (real) eigenfunctions form a checker-board pattern,
and their number $\nu_n$ is proportional to the product of the
angular and the ``surface" quantum numbers. Arranging the wave
functions by increasing values of the Laplace-Beltrami spectrum, we
obtain the nodal sequence, whose statistical properties we study. In
particular we investigate the distribution of the normalized counts
$\frac{\nu_n}{n}$ for sequences of eigenfunctions with $K \le n\le K
+ \Delta K$ where  $K,\Delta K \in \mathbb{N}$. We show that the
distribution approaches a limit as $K,\Delta K\rightarrow\infty$
(the classical limit), and study the leading corrections in the
semi-classical limit. With this information, we derive the central
result of this work: the nodal sequence of a mirror-symmetric
surface is sufficient to uniquely determine its shape (modulo
scaling).
\end{abstract}

\section{Introduction}
\label{sec:introduction} Nodal domains of a real, continuous
function are the maximally connected domains where the function does
not change its sign. The nodal domains of eigenfunctions of the
Laplacian on compact domains have been studied since Chladni first
observed the nodal structure of vibration modes of thin plates
(eigenfunctions of the bi-harmonic operator), in the early years of
the 19$^{{\rm th}}$ century. In the present manuscript we are not
interested in the geometric properties of nodal domains of Laplacian
eigenfunctions, but rather in their count.

Following Courant, we order the eigenfunctions so that the
corresponding eigenvalues form a non-decreasing sequence. Denoting
by $\nu_n$ the number of nodal domains of the $n$-th eigenfunction,
we form the normalized nodal sequence
$\xi_n:=\frac{\nu_n}{n},\textrm{ }n\in\mathbb{N}$. Courant's theorem
\cite{courant} guarantees that $\xi_n\leq 1$, and we would like to
study the distribution of the values of $\xi_n$ in the unit interval
$(0,1]$. In previous papers \cite{BGS,SUS} the distribution of the
$\xi_n$ for various planar domains and $2$-manifolds were studied,
and it was concluded that the features of the distribution depend
crucially on the type of classical dynamics it supports. If the
classical dynamics on the manifold (geodesics) are integrable (and
quantum mechanically separable - for such systems, actually,
quantum separablilty is equivalent to classical separability
\cite{havas}), the limit distribution exists, and displays certain
features which are common to all such systems. On the other hand, if
the classical dynamics is chaotic, the distribution of the
normalized nodal sequence is well reproduced by using a random wave
model for the eigenfunctions \cite {Berry77}. Bogomolny and Schmit
computed the mean and the distribution by using ideas from
percolation theory \cite {bogoschmidt}.

The first work with implications on the geometric content of the
nodal sequence was of Smilansky and Sankaranarayanan \cite {SUS},
where it was shown that the aspect ratio of a rectangular domain on
the plane (with Dirichlet boundary conditions) can be determined by
counting its nodal domains. In  \cite{GKS} the nodal sequence for
eigenfunctions of the Laplace-Beltrami operator for ``simple"
surfaces of revolution was discussed. A trace formula for the nodal
count was derived, and was shown to depend explicitly both on some
mean geometric properties of the surface, as well as the lengths of
its geodesics. In spite of the formal similarity between the
\emph{spectral} and the \emph{nodal} trace formulae, the geometrical
information is included in different ways. Further studies
\cite{GSS,BKP,Rami} have shown that isospectral domains have
different nodal sequences, thus supporting the conjecture that the
geometrical information is stored in the nodal and spectral
sequences in different ways. In the present work we go one step
further and inquire whether  \textit{one can   deduce the shape} (up
to scaling) \textit{of a domain given the distribution of the
normalized number of nodal domains,} or, paraphrasing the classical
spectral inversion question posed by M. Kac \cite{kac}, \textit{can
one count the shape of a drum?} It must be stressed that the only
use of the spectral information is lexicographical - it is ordered
as a non-decreasing sequence. Otherwise, there is no reference to
the actual values of the eigenvalues.

In the present manuscript we shall confine ourselves to the
integrable (and separable) case, particularly to a special class of
surfaces of revolution. We present new results which pertain to the
distribution $P(\xi,I_K)$ of the normalized nodal counts of
eigenfunctions with indices $n$ in the interval $I_K=[K,K+\Delta
K]$. In [1] it was showed that there exists a limit distribution
$P(\xi)$ when the size of the index interval $I_K$ becomes infinite
(corresponding to the semi-classical limit, $K\rightarrow\infty$).
We provide the leading order term of the difference between
$P(\xi,I_K)$ and the limit distribution $P(\xi)$:
\begin{equation}
P(\xi,I_K) = P(\xi) +\frac{1}{\sqrt{K}} P_1(\xi) + O\left (
\frac{1}{K} \right ) \ .
\end{equation}
We show that the knowledge of the function $P(\xi)$ and $P_1(\xi)$
suffices for nodal domain inversion, provided that the surface is
mirror-symmetric. In other words,  given the normalized nodal
sequence, we can deduce uniquely the profile function of the surface
of revolution (provided it is smooth and symmetric). Numerical
simulations were carried out for ellipsoids of revolution, which
illustrate our theoretical findings.

In what follows we shall use the classical notation of asymptotic analysis;
the standard `$O,o$' order notation, the symbol `$\sim$' standing for an asymptotic
relation, the symbol `$\asymp$' denoting same order of magnitude, and the symbol
`$\gg$' denoting greater order of magnitude.

\section{Surfaces of revolution}
\label{sec:surfaces} We consider surfaces of revolution
$\mathcal{M}$ in $\mathbb{R}^3$, which are generated by the complete
rotation of the line $y=f(x), \  x\in I:= [-1,1]$, about the $x$
axis. We confine our attention to a special subset of functions
which satisfy the following requirements:

\noindent \emph{i.} $ f^2$ is analytic in $I$, and vanishes at $\pm
1$, where $f(x) \sim a_{\pm}(1 \mp x)^{1/2}$, with $a_{\pm}>0$. This
requirement guarantees that $\mathcal{M}$ is compact, has no
boundary and  is smooth even at the points where it is intersected
by the axis of rotation.

\noindent \emph{ii.} The second derivative of $f$ is strictly
negative, so that $f(x)$ has a single maximum at some $x=x_{max}$,
where it reaches the value $f_{max}$. This requirement guarantees
convexity of $\mathcal{M}$.

Surfaces which satisfy the requirements above will be referred to as
\textit{simple surfaces of revolution}, and are convex, mild
deformations of ellipsoids of revolution. The induced Riemannian
metric on $\mathcal{M}$ is
 \begin{equation}
  {\rm d}s^2 = (1+f'(x)^2){\rm d}x^2 + f(x)^2 {\rm d}\theta^2 \ ,
  \label{eq:metric}
 \end{equation}
where the prime denotes differentiation with respect to $x$, and
$\theta\in[0,2\pi)$ is the azimuthal angle.

In the proceeding subsections we shall review the properties of
geodesics (classical mechanics) and the spectrum of the
Laplace-Beltrami operator (quantum mechanics) on $\mathcal{M}$.

\subsection{The geodesics}
 \label{subsec:geo}
The geodesics on $\mathcal{M}$ are the classical trajectories of
free motion. They can be derived from the Euler-Lagrange variation
principle with the Lagrangian
\begin{equation}
L=\frac{1}{2}\left ((1+f'(x)^2)\dot{x}^2 + f(x)^2
\dot{\theta}^2\right ) \ ,
 \label{eq:Lagrangian}
\end{equation}
where a dot above denotes time derivative. The angular momentum
along the axis of rotation $f(x)^2\dot{\theta}$ is conserved, and we
shall denote its value by $m$. The  momentum conjugate to $x$ is
$p_x=(1+f'(x)^2)\dot{x}$, and the conserved energy is
\begin{equation}
E=(1+f'(x)^2)\dot{x}^2 +\frac{m^2}{f(x)^2} \ .
 \label{eq:energy}
\end{equation}
It is convenient to introduce the action variable $n$,
\begin{equation}
\hspace{-1cm} n(E,m) := \frac{1}{2\pi} \oint p_x {\rm d}x =
\frac{1}{\pi}\int_{x_{-}}^{x_{+}}\sqrt{E f(x)^2-m^2 }
\frac{\sqrt{1+f'(x)^2}}{f(x)}{\rm d}x\ .
 \label{eq:action}
\end{equation}
Here, $x_{\pm}$ are the classical turning points, where $E f(x)^2
-m^2 =0$, with $x_{-}\le x_{max}\le x_{+}$, which correspond to two
meridians $\gamma_{\pm}$ (projections of caustics onto
$\mathcal{M}$) between which all geodesics with $m\neq 0$ wind
around $\mathcal{M}$. Real classical trajectories exist only if $E
>(m/f_{max})^2$. The convexity of $\mathcal{M}$ guarantees that the action variables
$(n,m)$ along with their conjugate angle variables constitute a
global coordinate system on phase space \cite{Zelditch98}.

The classical Hamiltonian $H(n,m)$ in the action-angle
representation is obtained by inverting (\ref {eq:action}) to
express  the energy in terms of $n$ and $m$.   $H(n,m)$ is a
homogeneous function of order $2$, i.e. $H(\lambda n,\lambda
m)=\lambda^2H(n,m),\textrm{ }\lambda> 0$ \cite {bleher94}. It
suffices, therefore, to study the function $n(m):= n(1,m)$, which
defines a smooth line $\Gamma$ in the $(n,m)$-plane (the projection
of the unit energy shell on the action plane). The function $n(m)$
is one of the main building blocks of the semi-classical theory
which will be used throughout this work. We shall list some of its
properties which will be used in the sequel:

\noindent \emph{1.} The reflection symmetry, $n(-m) =n(m)$, follows
from the definition (\ref{eq:action}). Thus, we restrict our
attention to $m\geq 0$ when referring to $\Gamma$.

\noindent \emph{2.} $n( |m| )$ is defined in the interval $I_{\mu}=
(0,m_{max}]$, where $ m _{max} = f_{max}$. In this interval, $n(m)$
is analytic and decreases monotonically since
\begin{equation}
\frac{{\rm d}n(m)}{{\rm d} m} = -\frac{m}{\pi}
\int_{x_{-}}^{x_{+}}\frac {1}{\sqrt{f(x)^2-m^2}
} \frac{\sqrt{1+f'(x)^2}}{f(x)}\ {\rm d}x\
\le 0 .
 \label{eq:der1}
\end{equation}

\noindent \emph{3.}  $n(m)$  assumes its maximum value  at $m=0$ ,
\begin{equation}
n(0) = \frac{1}{\pi} \int_{-1}^1\sqrt {1+f'(x)^2}\ {\rm d}x =
\frac{\mathcal{L}}{\pi} ,
 \label{eq:nof0}
\end{equation}
where $\mathcal{L}$ is the length of the rotating line. We show in
\ref {appendix1} that $n(m)$ is not analytic at $m=0$, and in that
vicinity
 \begin{equation}
  n(m) \sim \frac{\mathcal{L}}{\pi} - |m| \ .
  \label{eq:munear0}
 \end{equation}
At the other endpoint, $n(m)$ vanishes,
\begin{equation}
n(m) \sim   \  \sqrt{\frac{2} {\omega}} (m_{max}-m)\ \ , \ \
\omega =|2m_{max} f''(x_{max})|.
 \label{eq:munearmax}
\end{equation}

\noindent \emph{4.} The phase space volume is
\begin{equation}
\frac{1}{(2\pi)^2}\int_{\mathbb{R}^2_+} \Theta\Big(E-H(n,m)\Big){\rm d}v  =2 E
\int_0^{m_{max}}n(m){\rm d}m = 2  \mathcal{A} E \ . \label{eq:defA}
 \end{equation}
$\mathcal{A}$ is the area enclosed between the line $\Gamma$ and the
$n$ and $m$ axes. It is related to the area of $\mathcal{M}$ by
$||\mathcal{M}|| = 8\pi \mathcal{A}$.

\noindent \emph{5.} The computation of the higher derivatives of
$n(m)$ cannot proceed simply by taking the derivatives of (\ref
{eq:der1}) - the resulting integrals diverge. To overcome this
difficulty the integral defining $n(m)$ needs regularization
\cite{Gurarie}. This is done in \ref{appendix1}.

\noindent \emph{6.} Some authors (e.g., \cite{bleher94}) prefer to
use the Clairaut integral $\mathcal{I}$ instead of the angular
momentum. They are related by
\begin{equation}
\mathcal{I}= \frac{m}{\sqrt{2E}} \ . \label{eq:clair}
\end{equation}
The \emph{twist condition} is introduced in \cite{bleher94} to
distinguish the class of simple surfaces of revolution, for which
the dynamics are particularly simple. In the present notation, the
twist condition is expressed by the requirement
\begin{equation}
\frac{\partial^2  n(E,m) }{\partial m^2} \ne 0\ , \ \ \ \ \ 0\
<|m|\le m_{max}
 \label{eq:twist}
\end{equation}
i.e. $n(m)$ is either convex or concave on $(0,m_{max})$.

Throughout this paper we shall use the ellipsoid of revolution to
illustrate graphically our findings. The ellipsoids are generated by
$f(x)^2=\varepsilon^2(1-x^2)$, with $\varepsilon>0$ being the
eccentricity of the generating semi-ellipse.

The action variable $n(m)$ reduces to
\begin{equation}
n(m)=\frac{2x_{\pm}^2}{\pi}\int_0^1\frac{\sqrt{1-t^2}}
{1-x_{\pm}^2t^2}\sqrt{1-(1-\varepsilon^2)x_{\pm}^2t^2}{\rm d}t \ ,
\end{equation}
where $x_{\pm}=\pm \sqrt{1-m^2/\varepsilon^2}$ are the classical
turning points. This integral can be evaluated in terms of elliptic
functions giving,
\begin{equation}
\hspace{-20mm}
 n(m)=\frac{2}{\pi}\frac{1}{\varepsilon \sqrt{b}
}\Big[b{\rm E}(1-\frac{\varepsilon^2}{b})- (1-\varepsilon^2)m^2{\rm
K}(1-\frac{\varepsilon^2}{b})-
\varepsilon^4\Pi(1-\frac{\varepsilon^2}{m^2},1-\frac{\varepsilon^2}{b})\Big]
\ ,
\end{equation}
where $b=\varepsilon^4+(1-\varepsilon^2)m^2$, and
$${\rm
K}(k):=\int_0^{\pi/2}(1-k\sin^2\theta)^{-1/2}{\rm d}\theta,$$ $${\rm
E}(k):=\int_0^{\pi/2}(1-k\sin^2\theta)^{1/2}{\rm d}\theta$$
$$\Pi(k,l):=\int_0^{\pi/2}(1-k\sin^2\theta)^{-1}(1-l\sin^2\theta)^{-1/2}{\rm
d}\theta$$ are the complete elliptic integrals of first, second and
third kind respectively.

Figure 1. shows the functions $\Gamma:= n(m)$ for a few ellipsoids
of revolution

 \vspace{5mm}
\begin{figure}[h]
\centering \label{fig:picture1}
\includegraphics[width=4in]{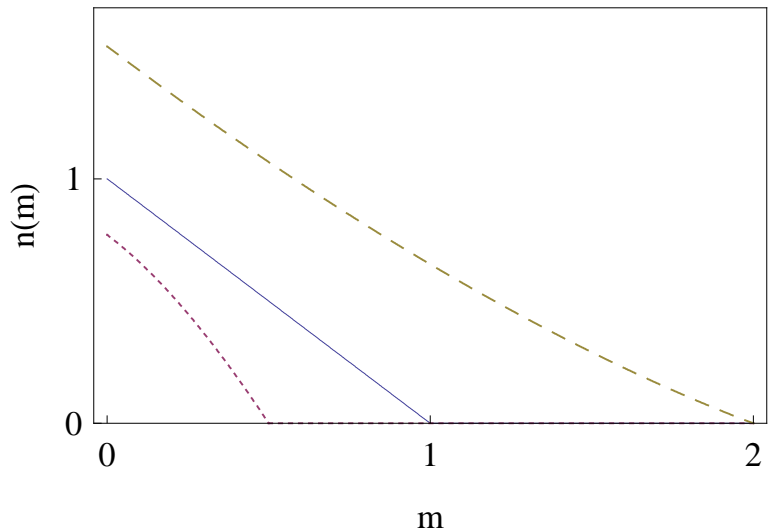}
\caption{\label{Figure.1} The curves $\Gamma$ for ellipsoids of
revolution with eccentricities $\varepsilon=0.5$ (prolate),
$\varepsilon=1$ (a sphere) and $\varepsilon=2$ (oblate) are shown as
dashed line, solid line and sparsely dashed line resp. }
\end{figure}

\subsection{The Laplace-Beltrami operator}
The Laplace-Beltrami operator on $\mathcal{M}$ reads
\begin{equation}
\Delta \ = \ -\frac{1}{f(x)\sigma(x)}\frac{ \partial \ }{ \partial
x}\ \frac{f(x)}{\sigma(x)}\frac{ \partial \ }{ \partial x} -
\frac{1}{f(x)^2}\frac{ \partial ^2\ }{ \partial \theta^2}\ ,
 \label{eq:LapBel}
\end{equation}
where $\sigma(x):= \sqrt {1+f'(x)^2}$. The domain of $\Delta$ are
$\Psi\in W_2^2(I \times \mathbb{S}^1)$, required to be
$2\pi$-periodic in $\theta$. Under these conditions, the operator is
self-adjoint, and its spectrum is discrete and non-negative. $\Delta
$ is separable, and the eigenfunctions can be written in the
product form $\Psi(x,\theta) = \exp(i m \theta)\ \psi_m(x)$, where
$m\in \mathbb{Z}$. It is convenient to introduce a new variable $t$,
through
\begin{equation}
{\rm d}t = \frac{\sigma(x)}{f(x)} {\rm d}x \ ,
 \label{eq:newvar}
\end{equation}
which maps the interval $[-1,1]$ to $\mathbb{R}\cup\{\infty\}$. For any $m$ and
eigenvalue $E_{n,m}$, the spectral equation for (\ref {eq:LapBel})
reduces to the Sturm-Liouville ODE
\begin{equation}
\left [- \frac{ {\rm d}^2\ }{{\rm d}t^2} +\left (\ m^2-E_{n,m}
f(x(t))^2\ \right) \right ]\psi_{n,m}(t) =0 \ .
 \label{eq:ode}
\end{equation}
The spectrum of the Laplace-Beltrami operator is doubly degenerate
for all $m \ne 0$, with $E_{n,-m}=E_{n,m}$. The semi-classical
spectrum is constructed by using the Einstein-Brillouin-Keller
approximation \cite {bleher94},
 \begin{equation}
E_{n,m}^{scl} = H(n+\frac{1}{2},m),\ n \in \mathbb{N}_0,\ m\in
\mathbb{Z}\ ,
 \label{eq:BSspectrum}
 \end{equation}
where $H(n,m)$ is the classical Hamiltonian defined in terms of the
action variables. The semi-classical approximation for the spectral
sequence with $m=0$ assumes a very simple form. Since
\begin{equation}
n+\frac{1}{2} = \frac{\sqrt{E}}{\pi}\int_{-1}^{1}
\sqrt{1+f'(x)^2}{\rm d}x = \frac{\sqrt{E}}{\pi}\mathcal{L} \ ,
\end{equation}
the semi-classical quantization condition reads:
\begin{equation}
E_{n,0}^{scl} =  \Big[\frac{\pi
(n+\frac{1}{2})}{\mathcal{L}}\Big]^2,\ n \in \mathbb{N}_0 \ .
 \label{eq:mzerospect}
\end{equation}

Because of the degeneracy of the spectrum, we have to choose a
particular representation of the wave functions. We do this by
associating $\cos(m\theta)$ with $m\ge 0$ and $\sin(m\theta)$ for
$m<0$, i.e.
\begin{eqnarray}
  \Psi_{n,m}(x,\theta) =\psi_{n,m}(x)\left \{
  \begin{array}{ll}
   \sin (m\theta) &  {\rm if} \ m<0    \\
   \cos (m\theta) &  {\rm if} \ m\geq 0 \ .
  \end{array}
  \right .
\end{eqnarray}

The nodal pattern of $\Psi$ is that of a checkerboard, typical for
separable systems (as a matter of fact, the nodal pattern remains a
checkerboard for any linear combination of the basis functions. It
is only rotated around the symmetry axis of the surface). For $m=0$,
the number of nodal domains is $\nu_{n,0}=n+1$, and for all other
$m$, $\nu_{n,m}= 2(n+1) |m| $. In summary
\begin{equation}
\nu_{n,m} = (n+1)(2 |m| + \delta_{m,0})\ \ .
 \label{eq:nu}
\end{equation}

To end this section we illustrate its content by an application to
the simplest surface - the sphere, considered here as a surface of
revolution with $f(x)^2= 1-x^2$.

The action variable (\ref {eq:action}) can be computed explicitly
\begin{equation}
n(E,m)=\frac{2\sqrt{E}}{\pi}\int_0^{\sqrt{1-m^2/E}}\frac{\sqrt{1-m^2/E-x^2}}
{1-x^2}\ {\rm d}x =\sqrt{E}\  -\ |m|\ .
 \label{eq:actsphere}
\end{equation}
Thus, $H(n,m) = (n+|m|)^2$, and the EBK quantization for the
spectrum is $E_{n,m} = (l+\frac {1}{2})^2 \sim l(l+1)$ where $l :=
n+|m|$. The $ (2l+1) $-fold degeneracy follows immediately by
counting the number of integer pairs $(n,m)$ which satisfy
$l=n+|m|$.

Turning to the quantum description, the variable $t$ defined in
(\ref{eq:newvar}) can be explicitly computed, $x=\tanh t$. Writing
$E=l(l+1), \ l\in \mathbb{N}_0$, transforms (\ref {eq:ode}) to
\begin{equation}
\left [- \frac{ {\rm d}^2\ }{{\rm d}t^2} +\left (\ m^2-
\frac{l(l+1)}{\cosh^2 t} \right) \right ]\psi_{l,m}(t) =0 \ ,
 \label{eq:odesphere}
\end{equation}
which is equivalent to the Legendre equation. Finally, the number of
nodal domains of spherical harmonics is known, and coincides with
(\ref {eq:nu}) when the identification $l = n+|m|$ is made.

\section{Counting nodal domains}
\label{counting} We shall start this section by reviewing some of
the general definitions and results obtained in \cite{BGS}, where
the limit distribution of the normalized nodal counts was first
studied. We shall then derive the next to leading term, and show
that it provides further information on the geometry of
$\mathcal{M}$.

The nodal structure of the wave functions was reviewed in the
preceding section, and an explicit expression for the dependence of
the number of nodal domains $\nu_{n,m}$ on the quantum numbers
$(n,m)$ is given in (\ref {eq:nu}).

The object which is investigated in this work is the \emph{nodal
sequence} which is defined as follows: Arrange the spectrum as a
non-decreasing sequence. This amounts to assigning to each pair of
quantum numbers $(n,m)$ a counting index $\mathcal{N}(n,m)$, which
gives the number of eigenvalues of the Laplace-Beltrami operator
(counted with multiplicity) which are \emph{strictly} smaller than
the eigenvalue $E_{n,m}$, i.e. $\mathcal{N}(n,m):=\#\{E\in{\rm
Spec}(\Delta):E < E_{n,m}\}$. Obviously
$\mathcal{N}(n,m)=N(E_{n,m})$, where $N(E)$ is the spectral counting
function.

To account for the spectral degeneracy, we modify the definition
above for  $|m|\ne 0$, so that $\mathcal{N}(n,|m|)
=\mathcal{N}(n,-|m|)+1$. The nodal sequence is the sequence of nodal
counts ordered by $\mathcal{N}$:
$\{\nu_{\mathcal{N}}\}_{\mathcal{N}=1}^{\infty}$. By this
convention, the systematic degeneracy of the spectrum is taken care
of.  In general, however, accidental degeneracies cannot be
excluded. The ordering ambiguity may appear, e.g.,  when the
degeneracy class involves states with different (non-negative) $m$
values, such as e.g., for the sphere. In this case a possible way to
remove this problem is to consider the sphere as a limiting case of
an ellipsoid with (a positive) eccentricity approaching zero. The
$m$ degeneracy is removed for any ellipsoid, and the order of the
eigenvalues is monotonic in $|m|$ for arbitrary small eccentricity.
Similar constructions can be used for other accidental degeneracies.

Courant's theorem \cite{courant} ensures that, for any ordering of the
eigenfunctions in the degeneracy classes
\begin{equation}
  \nu_{\mathcal{N}(n,m)} \le  \mathcal{N}(n,m)  \ .
\label{eq:courant}
\end{equation}

It is natural therefore,  to define the \textit{normalized nodal sequence}
\begin{equation}
\xi_{\mathcal{N}}\  :=\  {\nu_{\mathcal{N}}\over {\mathcal{N}} }\ \
, \ \ \ \ 0<\xi_{\mathcal{N}} \le 1 \ . \label{eq:nndom}
\end{equation}

We study the distribution of the values of the normalized nodal
sequence for a finite index set, $\mathcal{N}\in\{K,...,K+\Delta
K\}$,
\begin{equation}
 P(\xi,I_K )  :={1\over
\Delta
K}\sum_{(n,m)\in\mathbb{N}_0\times\mathbb{Z}}
\chi_{I_K}\left(\mathcal{N}(n,m)\right)\delta
(\xi - {\xi_{\mathcal{N}(n,m)}}) \ ,
 \label{eq:xidis}
\end{equation}
 and $\chi_{I_K}$ the characteristic function of the interval $I_K$,
 \begin{eqnarray}
  \chi_{I_K}(x) =\left \{
  \begin{array}{ll}
   1 &  {\rm if} \ K\le x\le K+\Delta K \    \\
   0 &  {\rm otherwise} \ .
  \end{array}
  \right .
 \end{eqnarray}

Since $P(\xi,I_K)$ is not a function but a distribution over the
interval $(0,1)$, one must take care in its manipulations. Most of
the limits and estimates are considered in the weak sense (e.g.
\cite{Jens}). In some sections though, related distributions are
viewed as functions (or to be more precise, the functions whose
values these distributions take on some subinterval of $(0,1)$). In
other cases they will be manipulated as functions after an
appropriate regularization.

In \cite {BGS}, it was assumed for convenience that $I_K$ grows
linearly in $K$, $\Delta K=gK$, $g$ being a positive constant (such
that $gK\in\mathbb{N}$). The existence of the limiting distribution
$P(\xi)$ of $P(\xi,I_K)$ in the $K \rightarrow \infty $ limit (the
classical limit) was proved, and its universal features were
presented. The existence can be proven by showing
\begin{equation}
\frac{1}{\Delta K}\sum_{j\in I_K}\varphi(\xi_j)
\longrightarrow\int_0^1P(\xi)\varphi(\xi){\rm d}\xi \ ,
\label{eq:limitdis}
\end{equation}
uniformly in $g$, for any smooth and compactly supported test
function $\varphi$ on $(0,1)$. Here, we shall repeat the derivation in more detail, and
also compute the difference between $P(\xi,I_K)$ and $P(\xi)$ to
leading order in $\frac {1}{\sqrt{K}}$ (again, for the linear case
$\Delta K=gK$ - the superlinear case $\Delta K\gg K$, follows
trivially).

We rewrite (\ref {eq:xidis}) as a sum of an isotropic component, to
which only isotropic states (with $m=0$) contribute, and an
anisotropic component (with $m\neq 0$),
\begin{eqnarray}
\label{eq:pixim}
P(\xi,I_K ) &=& P_{m=0}(\xi,I_K ) + P_{m \ne 0}(\xi,I_K ) \\
P_{m=0}(\xi,I_K )&=& {1\over g
K}\sum_{n}\chi_{I_K}\left(\mathcal{N}(n,0)\right)\delta \Big(\xi -
 \frac{n+1}{\mathcal{N}(n,0)}\Big) \nonumber \\
P_{m \ne 0}(\xi,I_K )&=& {1\over g K}\sum_{n,m\ne
0}\chi_{I_K}\left(\mathcal{N}(n,m)\right)\delta \Big(\xi -
\frac{2(n+1)|m|}{\mathcal{N}(n,m)}\Big)\ . \nonumber
\end{eqnarray}
Since we are interested in the semi-classical limit,  we are allowed
to make the following approximate steps, which incur errors of order
higher than $O(\frac{1}{\sqrt K})$.

 \noindent{\it i.} The spectrum is approximated by
$E_{n,m}\sim E^{scl}_{n,m}=H(n+\frac{1}{2},m)$, as
$m^2+n^2\rightarrow \infty$, which introduces a relative error
bounded by $O(\frac{1}{E_{n,m}})$ \cite {CDV}. In particular, as was
shown in (\ref{eq:mzerospect}), $E^{scl}_{n,0} = [\frac {\pi
(n+\frac{1}{2})}{\mathcal{L}}]^2$ .

\noindent {\it ii.} To the same order, ${\mathcal {N}}(n,m)$ can be
replaced by the first term in the Weyl series \cite {bleher94},
 \begin{equation}
  {\mathcal {N}}(n,m)  =  2 \mathcal {A}
H(n+\frac{1}{2},m) \Big(1 + O({1\over  E_{n,m}^{\frac{3}{4}}})\Big)\
,
 \label{eq:weil}
 \end{equation}
where ${\mathcal {A}}$ was defined in (\ref {eq:defA}).

Introducing these approximations in (\ref{eq:pixim}), we find that
$P_{m=0}(\xi,I_K )$ is $O(\frac{1}{K} )$ (in the weak sense) and
therefore we defer its computation to a later stage. The sums over
$(n,m)$ in (\ref{eq:pixim}) are computed using the Poisson summation
formula, decomposing $P_{m\neq0}(\xi,I_K)$ in a smooth and an
oscillatory part,
\begin{equation}
\hspace{-25mm}P_{m \ne 0}(\xi,I_K)=\bar P(\xi,I_K)+Q(\xi,I_K):=\bar
P(\xi,I_K)+\sum_{(N,M)\in
\mathbb{Z}^2\backslash\{0\}}Q_{N,M}(\xi,I_K) \ , \label{eq:poisson}
\end{equation}
where the Fourier coefficients are
\begin{eqnarray}
\hspace{-25mm} & &Q_{N,M}(\xi,I_K) \sim \nonumber \\
\hspace{-25mm}& &\ \ \ \ {2\over g K}\int_{\frac{1}{2}}^{\infty}
{\rm d}m \int_{-\frac{1}{2}}^{\infty}  \ {\rm e}^{2\pi i
(Mm+Nn)} \chi_{I_K}\Big(2\mathcal{A}H(n+\frac{1}{2},m)\Big)\delta \Big(\xi -
\frac{2(n+1)m}{2\mathcal{A}H(n+\frac{1}{2},m)}\Big){\rm d}n \ ,
 \label{eq:piximne0}
\end{eqnarray}
and obviously $\bar P(\xi,I_K)=Q_{0,0}(\xi,I_K)$.

The leading term in the sum above is the smooth term $\bar
P(\xi,I_K)$ which we calculate first. The oscillatory terms are of
lower order in $\frac{1}{\sqrt{K}}$ and will be computed in a
separate subsection.

Proceeding with the smooth part, we shift the integration variable
$n\mapsto n+\frac{1}{2}$ and write the result as
\begin{equation}
\hspace{-20mm}   \bar P(\xi,I_K) \sim{2\over g
K}\int_{\frac{1}{2}}^{\infty} {\rm d}m \int_{0}^{\infty}
\ \chi_{I_K}\Big(2\mathcal A H(n,m)\Big)\delta \Big(\xi -
\frac{(n+\frac{1}{2})m}{\mathcal A H(n,m)}\Big){\rm d}n\ \ . \label{eq:pmne0}
\end{equation}
We change the integration variables $(n,m)\mapsto ({\mathcal E},s)$,
where ${\mathcal E} =H(n,m)$, and $s$ is defined through the
relations
\begin{eqnarray}
{\rm d}{\mathcal E} &=& \omega_n{\rm d}n + \omega_m {\rm d}m \ \
;\ \  \omega_n  =  \ \frac{\partial H(n,m)}{\partial\ n} \ , \
\omega_m =\frac{\partial
H(n,m)}{\partial \ m} \ , \nonumber \\
{\rm d}s &=&\frac{-\omega_m {\rm d}n + \omega_n {\rm
d}m}{\omega_n^2+\omega_m^2 } \ .
 \label{eq:transform}
\end{eqnarray}
Note that with this definition the Jacobian is unity, and ${\rm
d}n{\rm d}m={\rm d}{\mathcal E}{\rm d}s$. Thus, (\ref {eq:pmne0}) is
reduced to
\begin{equation}
\hspace{-25mm} \bar P(\xi,I_K )\sim{2\over g
K}\int_{\frac{K}{2\mathcal{A}}}^{\frac{K(1+g)}{2\mathcal{A}}}{\rm
d}{\mathcal E}\int_{\Gamma}  \ \Theta\Big(m(s)-\frac{1}{2\sqrt
\mathcal{E}}\Big) \delta \left (\xi - \frac{
n(s)m(s)}{\mathcal{A}}-\frac{m(s)}{2\sqrt \mathcal{E}\mathcal A
}\right ){\rm d}s \ ,\
 \label{eq:pmne0reduced}
\end{equation}
where $\Theta (x)$ is the Heavyside step function. The pair of
functions $\{n(s),m(s)\}$ constitute a parametric representation of
the line $\Gamma$, along which $H\Big(n(s),m(s)\Big) =1$. This
allows the scaling by $\sqrt \mathcal{E}$ which appears in
(\ref{eq:pmne0reduced}).

The expression above can be further reduced by the following
observation. On the line $\Gamma $ we have $\omega_n{\rm d}n +
\omega_m {\rm d}m =0$, which induces a symplectic structure (where
$n$ and $m$ play the roles of canonically conjugate variables and
$s$ is the ``time")
\begin{equation}
\frac{{\rm d}m(s)} {{\rm d} s} =  \omega_n   \ \ ; \ \ \frac{{\rm
d}n(s)} {{\rm d} s} =  - \omega_m \ .
 \label{eq:symp1}
\end{equation}
Another important identity follows from the fact that $H(n,m) $ is
homogeneous of order 2 in $(n,m)$. We can write $ n= n(H,m)=
\frac{1}{\sqrt{H}}n(1,\frac{m}{\sqrt{H}})$, from which we deduce
that on the line $\{H(n,m)=1\}$,
\begin{equation}
\frac{{\rm d}m(s)} {{\rm d} s} = \omega_n =\frac{2}{n(m)-m n'(m)}
\ ,
 \label{eq:symp2}
\end{equation}
where $ n'(m) := \frac{{\rm d}n(m)} {{\rm d} m}$ . Thus,
(\ref{eq:pmne0reduced}) takes the form
\begin{equation}
\hspace{-25mm} \bar P(\xi,I_K )\sim{1\over g
K}\int_{\frac{K}{2\mathcal{A}}}^{\frac{K(1+g)}{2\mathcal{A}}}{\rm
d}{\mathcal E}\int_{\frac{1}{2\sqrt \mathcal{E}}}^{m_{max}}
 \ \Big|n(m)-m n'(m)\Big| \delta \left (\xi - \frac{ n(m)m
}{\mathcal{A}}-\frac{m}{2\sqrt \mathcal{E}\mathcal A }\right ){\rm
d}m \ .\
 \label{eq:pmnmn}
\end{equation}
A similar transformation can also be applied in the computation of
the oscillatory integrals $Q_{N,M}(\xi,I_K )$.
\subsection{The limit distribution}
The limit distribution is the leading term in the expansion of the
integral (\ref{eq:pmne0reduced}) in powers  $O(\frac{1}{\sqrt K})$.
Taking the limit $K\rightarrow\infty$, the $\mathcal{E}$-integral
can be directly performed, resulting with
 \begin{equation}
P(\xi) = {1\over {\mathcal A}}\int_{\Gamma} \  \delta\left
( \xi-{m(s) n(s)\over {\mathcal A}}\right ){\rm d}s  \  .
\label{eq:xidist}
\end{equation}
This expression allows a simple geometrical interpretation: the
product $n(s) m(s)$ is the area of the rectangle whose vertices are
the points $(m(s),n(s))$ on $\Gamma$, its projections $(m(s),0)$ and
$(0,n(s))$ on the two axes and the origin (See Figure 2). $P(\xi)$
is the probability distribution of the areas of these rectangles
(scaled by ${\mathcal A}$, and therefore smaller than 1). The areas
of rectangles which are based on points which are either near the
$m$ or the $n$ axes approach $0$ . Since $\Gamma$ is either convex
or concave, there exists a unique point where $\xi$ reaches the
maximal value of scaled areas, $\xi_{max }$. Thus, $P(\xi)\equiv 0$
for $\xi \ge \xi_{max}$.

\begin{figure}[h]
\centering \label{fig:picture2}
\includegraphics[width=4in]{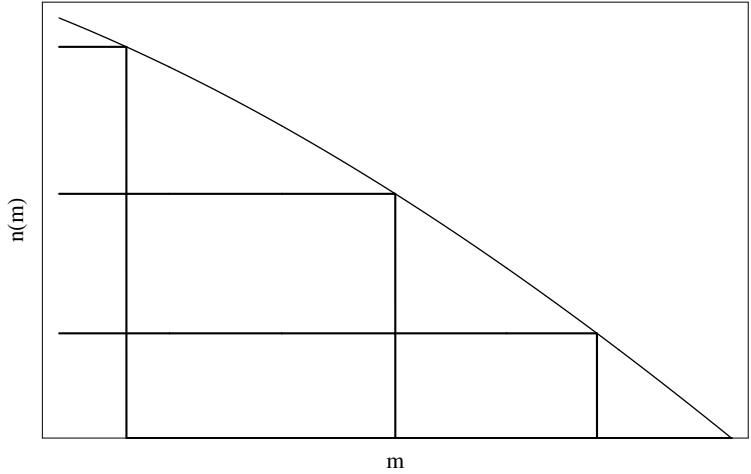}
\caption{\label{full}  The geometric interpretation of $P(\xi)$
(Computed for an ellipsoid of revolution with eccentricity
$\varepsilon=0.5$).}
\end{figure}

Changing integration variables using  (\ref{eq:symp2}),  we get
\begin{equation}
P(\xi) ={1\over {2 \mathcal A}}\int_0^{m_{max}}\delta\left (
\xi-{m\ n(m)\over {\mathcal A}}\right )\left|n(m)-m n'(m)\right|
{\rm d}m\ .
 \label {eq:symp3}
\end{equation}
Integrating (\ref{eq:symp3}) we get,
\begin{eqnarray}
P(\xi) =\left\{
\begin{array}{l}
\ \ \ \ \ \  0 \ \ \ \ \ \ \ \ \ \ \ \ \ \ \ \ \ \ \ \ \ \ \ \ \ \
\ \ \xi \
\ge\ \xi_{max} \ ,        \\  \\
 {1\over 2}\sum \left |{n(m)-mn'(m) \over n(m)+mn'(m)}
\right|_{\xi={ m n(m)\over{\mathcal A}}} \ \ \xi \ < \ \xi_{max} \ .
\end{array}
\right.
 \label {eq:integp}
\end{eqnarray}
The sum is over the real values of $m>0$  which satisfy $\xi={m
n(m)\over{\mathcal A}}$. In the vicinity of $\xi_{max}$, typically
two solutions coalesce, leading to a square root singularity of
$P(\xi)$ at that point. The vanishing of $P(\xi) $ in the interval
$[\xi_{max},1]$, and the square root singularity at $\xi_{max}$ are
the universal features which characterize the nodal domain
distributions for separable systems (in 2-d) in general, and simple
surfaces of revolution, in particular. Using (\ref {eq:symp3}), it
is easy to check normalization, $\int_0^{1} P(\xi){\rm d}\xi =1$.

The form (\ref {eq:integp}) for the $\xi$ distribution can be
further simplified since for simple surfaces of revolution the twist
condition (\ref {eq:twist}) is satisfied, and there are only two
values of $m$ which solve  $\xi={ m\ n(m)\over{\mathcal A}}$. Denote
them by $m_{-}(\xi)$  and $m_{+}(\xi)$ ($m_{-}(\xi) \le
m_{+}(\xi)$). They coalesce at $\xi=\xi_{max}$. The value of $m$
where the sole maximum of $mn(m)$ occurs is denoted by $m_0 =
m_{\pm}(\xi_{max})$. The  two functions $m_{+}(\xi)$ and
$m_{-}(\xi)$, together, provide a parametric representation of the
curve $\Gamma$, since $n_{\pm}(\xi)=\frac{{\mathcal
A}\xi}{m_{\pm}(\xi)}$ where $0 <\ \xi < \xi_{max}$. This
parametrization will be used often in the subsequent discussion.

To leading order, $m_{\pm}(\xi)\sim m_0(1\pm\sqrt{|\zeta|})$, where
$\zeta=\frac{\xi-\xi_{max}}{\xi_{max}}$, and so from
(\ref{eq:integp}) we deduce that, in the left neighborhood of
$\xi_{max}$
\begin{equation}
P(\xi)\sim \frac{1}{\sqrt{1-\xi/\xi_{max}}} \ .
\end{equation}

\begin{figure}[h]
\centering \label{fig:picture3}
\includegraphics[width=4in]{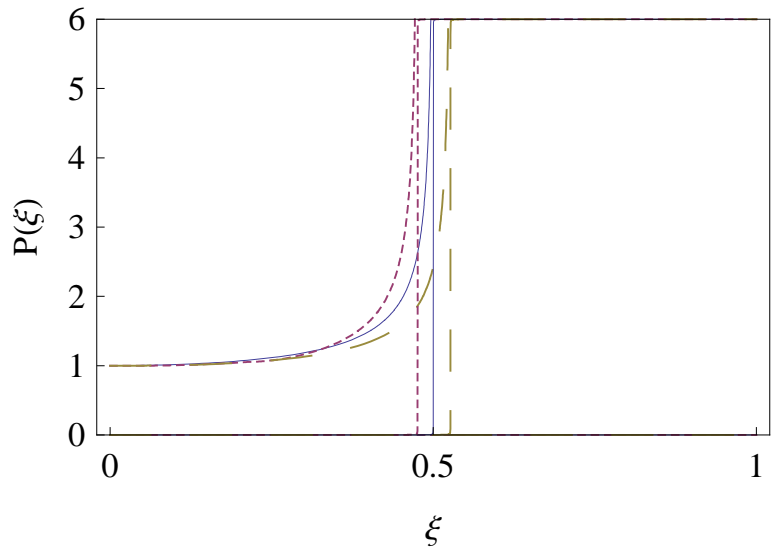}
\caption{\label{full}  $P(\xi)$ for ellipsoids of revolution with
eccentricities $\varepsilon=0.5,1$ and $2$ (sparsely dashed line,
solid line and dashed line resp.) }
\end{figure}

One can re-arrange (\ref{eq:integp}) to obtain another expression
for the limiting distribution. For $\xi<\xi_{max}$,
\begin{equation}
P(\xi) = \xi \frac{{\rm d}\ }{{\rm d}\xi} \log \frac
{m_{-}(\xi)}{m_{+}(\xi)} \ .
 \label{eq:explicit}
\end{equation}
This expression  is quite revealing, because it can be inverted to
provide the function $\frac {m_{-}(\xi)}{m_{+}(\xi)}$ based on
information derived only from the nodal sequence. We shall show
below that the next to leading expression in the expansion of
$P(\xi,I_K)$ provides another relation between $m_{-}(\xi)$  and
$m_{+}(\xi)$. Solving the two equation we obtain a complete
parametric representation of the curve $\Gamma$ (or $n(m)$). We
shall also show that $\Gamma$ defines uniquely the function $f(x)$,
when $f(x)$ is symmetric about $x=0$. This will prove our claim that
the nodal sequence for symmetric surfaces of revolution can be
inverted and provide the ``shape of the drum"!

The behavior of $P(\xi)$ near $\xi=0$ can be easily extracted using
(\ref{eq:explicit}). Since $\xi \rightarrow 0$ implies either $m\rightarrow 0$ or
$n(m)\rightarrow 0$ along $\Gamma$, we may use the linear approximations
for $n(m)$ as given by (\ref {eq:munear0}) and (\ref{eq:munearmax}),
respectively. The equation $\frac{m\ n(m)}{\mathcal{A}}=\xi$ reduces
to a quadratic equation, and its solutions define the two branches
$m_{-}(\xi)$ and $m_{+}(\xi)$ as
\begin{eqnarray}
\label{eq:nearxi0}
 m_{-}(\xi) &\sim&\
 \frac{\mathcal{L}}{2\pi}\ \left (1-\sqrt{1-2\xi\
\frac{2\mathcal{A}\pi^2}{\mathcal{L}^2}\ }\ \ \right
)\nonumber \\
 m_{+}(\xi) &\sim& \frac{m_{max}}{2}\left( 1+
  \sqrt{1- 2\xi \ \frac {\mathcal{A} \sqrt{2\omega}}
  {m_{max}^2 }}
  \ \right)\ .
\end{eqnarray}
Substituting in (\ref{eq:explicit}) we get
\begin{equation}
P(\xi) \sim \frac{1}{2} \left
(\frac{1}{\sqrt{1-\frac{2\xi}{\eta_{-}}}}
+\frac{1}{\sqrt{1-\frac{2\xi}{\eta_{+}}}} \right )\ ,
\label{eq:xismall}
\end{equation}
where,
 \begin{equation}
\hspace{-15mm} \eta_{-}=\frac{n(0)^2}{2\mathcal{A}}=
\frac{\mathcal{L}^2}{2\pi^2\mathcal{A}} \ \ ; \ \ \eta_{+} =
\frac{|n'(m_{max})|\
m_{max}^2}{2\mathcal{A}}=\frac{m_{max}^2}{\sqrt{2\omega}
\mathcal{A}}\ \ ;
 \ \  \eta_0= \frac{m_{max}^2}{2\mathcal{A}} \ .
 \label {eq:etadef}
\end{equation}
The relation (\ref {eq:xismall}) shows that $\lim_{\xi\rightarrow
0+}P(\xi)=1$, independently of the surface under consideration -
another universal feature to be added to the aforementioned ones.
Moreover, it shows that one can extract the dimensionless geometric
parameters $\eta_{-}$ and $\eta_{+}$ from $P(\xi)$ in the
neighborhood of $\xi= 0$. They are directly related to the
properties of the line which generates  $\mathcal{M}$ through its
length, maximum distance from the axis of revolution and its
curvature at the maximum. Note, however, that the nodal sequence is
composed of integers, and in contrast to the spectral sequence it is
invariant under isotropic scalings on $\mathcal{M}$. Therefore, only
dimensionless quantities can be extracted from it. Here, all the
lengths are expressed in units of $\sqrt{\mathcal{A}}$. (The
parameter $\eta_0$ in (\ref{eq:etadef}) is another dimensionless
parameter which we define here even though it will appear only
later).

The limit distributions for  three ellipsoids of revolution are
shown in Figure 3. Computing $\xi_{max}$ as a function of the
eccentricity reveals that $\xi_{max}(\varepsilon)$ is a
monotonically decreasing function which  varies between
$\xi_{max}(0) \approx .550$ and $\xi_{max}(\infty) \approx .464$ as
shown in Figure 4. Thus, one can deduce the eccentricity from the
nodal sequence just by determining the support of $P(\xi)$.

\begin{figure}[h]
\centering \label{fig:picture8}
\includegraphics[width=4in]{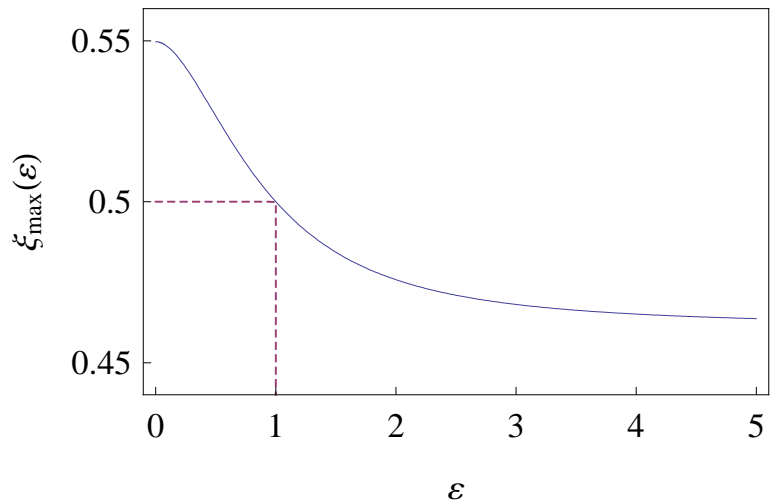}
\caption{\label{full}  $\xi_{max}$ as a monotonically decreasing
function of the eccentricity $\varepsilon$ for ellipsoids of
revolution. The dashed line marks $\varepsilon =1$ - the sphere -
for which $\xi_{max}=.5$ as in (\ref {eq:sphere p(xi)}).}
\end{figure}

As another application of (\ref{eq:explicit}), and as an
illustration, we derive $P(\xi)$ for the sphere. From (\ref
{eq:nsphere}) we get that for the sphere $n(m)=1-m$ which
immediately gives
\begin{equation}
P(\xi)= \frac{1}{\sqrt{1-2\xi}} \ ; \ \ \  0\ < \xi \ < \frac{1}{2}
\ . \label{eq:sphere p(xi)}
\end{equation}
The same result can also be obtained directly. The spectrum consists
of the values $E_{n,m}=n(n+1)$ which are independent of $m$ ($|m|\le
n$) and  are $(2n+1)$-fold degenerate. The eigenfunctions are the
spherical harmonics, and for the sake of counting their nodal
domains, we consider them in their separable basis. The number of
zeros of $P_n^m(\cos\theta)$ is $n$ when $m = 0$, and $n-|m|+2$
otherwise. The number of nodal domains are respectively $n+1$ and
$2|m|(n-|m|+1)$. The counting function for the $2n+1$ degenerate
states with a given $n$ satisfies
\begin{equation}
n^2 \le \mathcal{N}(n,m)< (n+1)^2 \ .
\end{equation}
The ordering within the set is arbitrary, but in the limit of large
$n$ it is immaterial, since to leading order  we can write
$\mathcal{N}(n,m)=n(n+1)(1 +O(\frac{1}{n}))$. Hence, the
contribution of the $n$-fold degenerate set to $P(\xi)$ is
\begin{eqnarray}
 P(\xi;n) &=& \frac{1}{n}\sum_{m=1}^n\delta
\Big(\xi-\frac{2m(n-m+1)}{n(n+1)}\Big) +O(\frac{1}{n})\nonumber \\
&\rightarrow& \ \int_0^1 \delta \Big(\xi -2\mu(1-\mu)\Big){\rm d}\mu\ =\
\frac{1}{\sqrt{1-2\xi}} \ .
\end{eqnarray}

\subsection{The next to leading terms}

Several terms contribute $O(\frac{1}{\sqrt{K}})$ corrections. In the
following we shall address them in detail.

\noindent \emph{The leading correction to $\bar P(\xi,I_K )$}.
Starting again from (\ref {eq:pmnmn}), we see that the
$m$-integration results in a sum of two terms, coming from the two
branches of solutions of $\xi=\frac {m\
n(m)}{\mathcal{A}}+\frac{m}{2\sqrt \mathcal{E}\mathcal
A}+O(\frac{1}{\mathcal{E}}) $ which constitute the two-point support
of the $\delta$ function in the integral. They differ by amount
$\delta m_{\pm}(\xi)$ of $O(\frac{1}{\sqrt \mathcal{E}})$ from the
values $m_{\pm}(\xi)$ which were introduced for the computation
(\ref{eq:explicit}) of the limit distribution,
\begin{equation}
\hspace{-20mm} \delta m_{\pm}(\xi) =-\frac{1}{2\sqrt{\mathcal{E}}}
\frac{m }{ m n'(m )+n(m ) }+O(\frac{1}{\mathcal{E}})\\ , \ {\rm \ computed\ at}\
m=m_{\pm}(\xi) \ . \label{eq:approxdel}
\end{equation}
As long as $(m +\delta m)_{\pm}$ remain inside the integration
range, one can proceed with the computation towards the
semi-classical asymptotic expansion

\begin{equation}
\bar P(I_K) = P +\frac{1}{\sqrt{K}}P_1+O(\frac{1}{K}) \ .
 \label{eq:I1next}
 \end{equation}
Of course, $P$ is the limit distribution (\ref {eq:integp}), and
\begin{equation}
P_1(\xi)=-\sqrt{\frac{\mathcal{A}}{2}}\frac{\sqrt{1+g}-1}{g}\Big(p_-(\xi)+p_+(\xi)\Big) \ ,
\label{eq:p1}
\end{equation}
where $p_{\pm}(\xi)$ are given by
\begin{equation}
\hspace {-25mm} p_{\pm}(\xi)=
\frac{\frac{1}{n(m)}}{1+m\frac{n'(m)}{n(m)}} \left
|\frac{1-m\frac{n'(m)}{n(m)}}{1+m\frac{n'(m)}{n(m)}}\right | \left
[-1+\frac{2 m\left (m\frac{n'(m)}{n(m)}\right
)'}{1-(m\frac{n'(m)}{n(m)})^2}\right ], \ {\rm \ computed\ at}\
m=m_{\pm}(\xi) \ .
 \label{eq:I1next1}
 \end{equation}
These expressions can be further simplified. To do so, we
differentiate $m_{\pm}(\xi)n(m_{\pm}(\xi))=\mathcal{A}\xi$, with
respect to $\xi$ and get
\begin{equation}
\hspace{-15mm} n=\frac{\mathcal{A}\xi}{m} \ \ ; \ \ \
n'=n\Big(\frac{1}{\xi\frac{{\rm d} m}{{\rm d}\xi}}-\frac{1}{m}\Big)\
\ ;
 \ \  n''=n\Big(\frac{2}{m^2}-\frac{2}{\xi m\frac{{\rm d} m}{{\rm d}\xi}}
 -\frac{\frac{{\rm d^2} m}{{\rm d}\xi^2}}{\xi ( \frac{{\rm d} m}{{\rm d}\xi} )^3}\Big) \ ,
\end{equation}
computed at $m=m_{\pm}(\xi)$. By substituting the above in the
defining expressions for $p_{\pm}(\xi)$, we obtain after some
straightforward calculations
\begin{equation}
p_{\pm}(\xi)=\pm\frac{1}{\mathcal{A}}\left( \frac{{\rm d}
m_{\pm}}{{\rm d}\xi}+2\xi\frac{{\rm d^2} m_{\pm}}{{\rm d}\xi^2}
\right) \ = \ \pm \frac{2}{\mathcal{A}}\sqrt{\xi}\ \frac{{\rm d}}{
{\rm d}\xi}\left( \sqrt{\xi}\ \frac{{\rm d} m_{\pm}}{{\rm
d}\xi}\right )\ .
 \label{eq:ppmexplicit}
\end{equation}
Hence,
\begin{equation}
P_1(\xi)=-\sqrt{\frac{2}{\mathcal{A}}}\frac{\sqrt{1+g}-1}{g}
\sqrt{\xi}\ \frac{{\rm d}}{ {\rm d}\xi}\left( \sqrt{\xi}\ \frac{{\rm
d} } {{\rm d}\xi}\left[m_{+}(\xi)-m_{-}(\xi) \right ]\right )\  \ .
\label{eq:p1inal}
\end{equation}
This is an explicit expression which provides the leading correction
in terms of the difference $m_{+}(\xi)-m_{-}(\xi)$ between the two
branches of the parametric representation of $\Gamma$. Together with
(\ref {eq:explicit}) it forms the basis for the inversion procedure
which will be discussed in detail in the next section.

The conditions for the validity of the above approximation are not
satisfied if $\xi$ is in the $O(\frac{1}{\sqrt{K}})$ neighborhood of
either $0$ or $\xi_{max}$. Near $\xi_{max}$ (\ref {eq:approxdel})
diverges, while near $\xi = 0$,  $(m +\delta m)_{\pm}$ may lie
outside of the integration range
$[\frac{1}{2\sqrt{\mathcal{E}}},m_{max}]$. To get the correct
expressions for $\bar P(\xi,I_K)$ in the vicinity of the extreme
values of $\xi $, we use several variations of the same trick:
within the problematic domains of integration, we approximate $n(m)$
as a linear function of $m$. The argument of the $\delta$ functions
become quadratic functions of $m$. The support of the $\delta$
functions can be evaluated explicitly, and the $m$-integrations can
be performed exactly. The remaining $\mathcal{E}$-integrations turn
out to be straightforward, so that explicit expressions for $\bar
P(\xi,I_K)$ in the vicinity of $\xi=0$ and $\xi=\xi_{max}$ are
obtained.

\textit{Behavior of $\bar P(\xi,I_K)$ in the neighborhood of
$\xi=\xi_{max}$}. We recall that $\xi_{max}$ is defined as the
maximum value of $\frac{m\ n(m)}{\mathcal{A}}$, which occurs at
$m_0$, where $n(m_0)+m_0\ n'(m_0)=0$. Thus, in the neighborhood  of
$m_0$ we can approximate $n(m)\sim n_0-\frac{n_0}{m_0}(m-m_0)$ where
$n_0=n(m_0)=\frac{\mathcal{A}\xi_{max}}{m_0}$ (the non-existence of
other critical points is guaranteed by the twist codition). With
this approximation, the argument of the $\delta$ function in (\ref
{eq:pmnmn}) is quadratic in $m$. The integrations over $m$ and
$\mathcal{E}$ have to be carried out with attention to the
requirement that the support of the $\delta$ remains within the
integration range. After some lengthy but straightforward
manipulations one gets,
\begin{equation}
\bar P(\xi,I_K)\sim \frac{2  }{g K \eta_{max}}
\int_{\sqrt{\frac{1}{(1+g)K\eta_{max}}}}^{\sqrt{\frac{ 1 }{K\eta_{max}}}}
\Theta \left (y-\frac{\zeta}{2}\right )\frac{y^{-3} {\rm
d}y}{\sqrt{y^2+2y-\zeta}}
\end{equation}
 and
 \begin{equation}
\eta_{max}=\frac{(4n_0)^2}{2\mathcal{A}} \ \  ;\ \ \ \zeta =
\frac{\xi-\xi_{max}}{\xi_{max}} \ .\label{:eqetazeta}
 \end{equation}
$\eta_{max}$ is another dimensionless parameter which characterizes
$\bar P(\xi,I_K)$ near the borders of its support.  Performing the
integral we get,
\begin{eqnarray}
\label{eq:ximexp}
 \hspace{-15mm} \bar P(\xi,I_K)&\sim&\frac{1}{g K\eta_{max}} \Big(
F(y^{\uparrow},\zeta)-F(y_{\downarrow},\zeta)\Big)\\
\hspace{-15mm} F(y,\zeta)
  &=&\left( \frac{1}{\zeta \
y^2}+\frac{3}{\zeta^2\
y}\right)\sqrt{y^2+2y-\zeta} \   \nonumber \\
\hspace{-15mm}&+& \left( \frac{1}{\zeta }+ \frac{3}{\zeta^2
}\right)\left \{
\begin{array} {ll}
\frac{-1}{\sqrt{|\zeta|}}{\rm
arctanh}\frac{|\zeta|+y}{\sqrt{|\zeta|} \sqrt{y^2+2y-\zeta} }
&\ \ \ \ \zeta\le 0 \\
\frac{1}{\sqrt{\zeta}}\arctan\frac{-\zeta+y}{\sqrt{ \zeta }
\sqrt{y^2+2y-\zeta} } &\ \ \ \ \zeta > 0
\end{array}
\right . \nonumber \\
\hspace{-15mm}& &y^{\uparrow}= \sqrt{\frac{1}{K\eta_{max}}} \ \ ;
 \ \ y_{\downarrow}=\ \ \ \ \ \ \sqrt{\frac{1}{(1+g)K\eta_{max}}}
 \ \ \ \ \ \ \  \ \ \ \    {\rm for}
\ \ \   \zeta \le 0 \nonumber\ , \\
\hspace{-15mm}& &y^{\uparrow}= \sqrt{\frac{1}{K\eta_{max}}} \ \ ; \ \
y_{\downarrow}= \max
\left\{\frac{\zeta}{2},\sqrt{\frac{1}{(1+g)K\eta_{max}}}\right\} \ \
{\rm for}\ \ \
 \zeta > 0 \ . \nonumber
\end{eqnarray}
This expression includes the corrections to the limit distribution
at its most noticeable feature, namely its singularity at
$\xi_{max}$. For finite $K$, the square-root singularity is replaced
by a continuous function which reaches beyond $\xi_{max}$, shifts
the maximum from $\xi_{max}$ to
$\xi_{max}(K,g)=\xi_{max}+2\sqrt{\frac{1}{(1+g)K\eta_{max}}}$ and
extends the support of $P(\xi,I_K)$ up to
$\xi_{max}+2\sqrt{\frac{1}{K\eta_{max}}}$. As $K \rightarrow
\infty$, the expression converges to the limit. The application of
the above theory for an ellipsoid  of revolution for two values of
$K$ are shown in Figures 5. and 6.

\begin{figure}[h]
\centering \label{fig:picture4}
\includegraphics[width=4in]{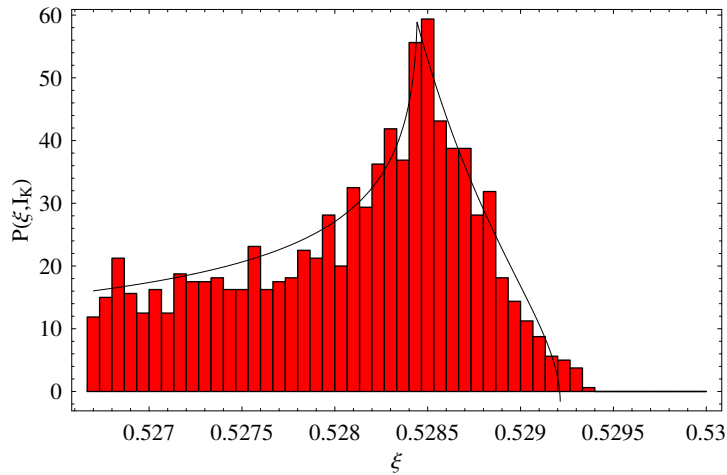}
\caption{\label{full}  Local behavior of $P(\xi,I_K)$ at
$\xi=\xi_{max}$ for an ellipsoid of revolution with eccentricity
$\varepsilon=0.5$ ($K=24000$ and $g=1$). }
\end{figure}

 \vspace{10mm}

\begin{figure}[h]
\centering \label{fig:picture5}
\includegraphics[width=4in]{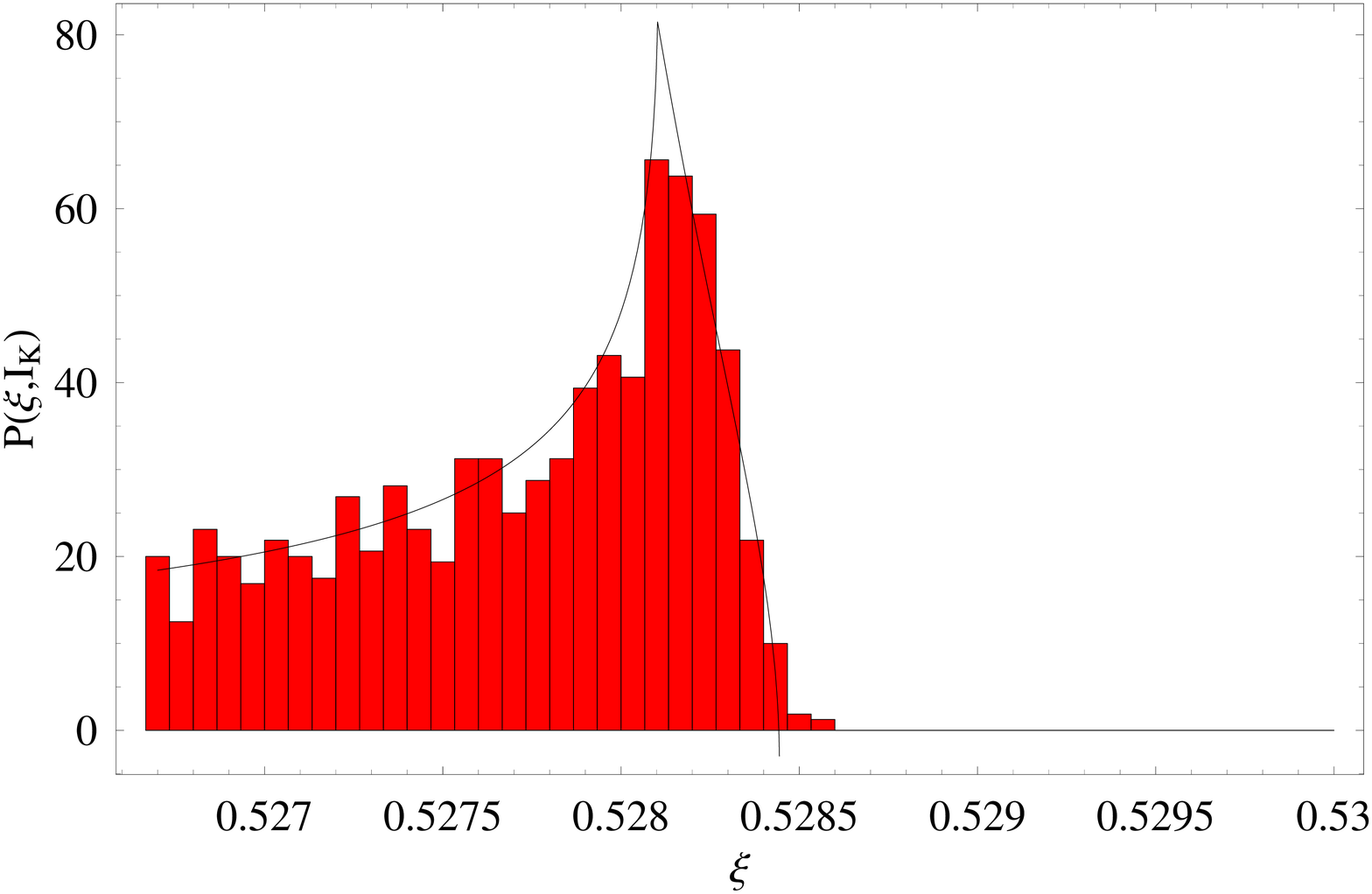}
\caption{\label{full}  Local behavior of $P(\xi,I_K)$ at
$\xi=\xi_{max}$ for an ellipsoid of revolution with eccentricity
$\varepsilon=0.5$ ($K=48000$ and $g=0.5$). }
\end{figure}

\textit{Behavior of $\bar P(\xi,I_K)$ in the neighborhood of
$\xi=0$}. In the neighborhood of $\xi=0$ we have contributions from
the support of the delta function  in (\ref{eq:pmnmn}) from the
neighborhood of $m=0$ - provided that $(m+\delta
m)_->\frac{1}{2\sqrt{\mathcal{E}}}$ - and $m=m_{max}$ - provided
that
$\xi>\frac{1}{\mathcal{A}}\Big(mn(m)+\frac{m}{2\sqrt{\mathcal{E}}}\Big)_{m_{max}}$.
Putting all contributions together, we have
\begin{eqnarray}
 \label {eq:pixinear0}
\hspace{-25mm}\bar P(\xi,I_K) \ \ \sim   \nonumber \\
\hspace{-25mm} \frac{1}{2}\left \{
\begin{array}{ll}
0 &  {\rm for}\ \ \ \ \ \  0\ \ \ \ \ < \xi \le \sqrt{\frac{\eta_{-}}{K(1+g)}}\\
\frac{1}{g}\frac{1}{1-\xi/\eta_-}\Big[(1+g)-\frac{\eta_-}
{K\xi^2}\Big] & {\rm for}  \ \sqrt{\frac{\eta_{-}}{K(1+g)}} < \xi \le \sqrt{\frac{\eta_-}{K}}\\
\frac{1}{1-\xi/\eta_-} & {\rm for}  \ \ \ \ \
\sqrt{\frac{\eta_{-}}{K}}\ \ < \xi
\nonumber
\end{array}\right .\\
\hspace{-25mm}+\frac{1}{2}\left \{
\begin{array}{ll}
0 &  {\rm for}\ \ \ \ \ \  0\ \ \ \ \ < \xi \le \sqrt{\frac{\eta_{0}}{K(1+g)}}\\
\frac{1}{g}\frac{1}{1-\xi/\eta_+}\Big[(1+g)-\frac{\eta_0}{K\xi^2}\Big]
& {\rm for}  \
\sqrt{\frac{\eta_{0}}{K(1+g)}} <
\xi \le \sqrt{\frac{\eta_0}{K}}\\
\frac{1}{1-\xi/\eta_+} & {\rm for}  \ \ \ \ \
\sqrt{\frac{\eta_{0}}{K}}\ \ <\xi \ .

\nonumber
\end{array}\right .\\
\end{eqnarray}
The most important feature in (\ref {eq:pixinear0}) is that it
shifts the support of $\bar P(\xi,I_K)$ away from $\xi=0$ to
$\min\Big\{\sqrt{\frac{\eta_-}{K(1+g)}},\sqrt{\frac{\eta_0}{K(1+g)}}\Big\}$.
Note that most of the expressions which make up $P(\xi,I_K)$ in the
neighborhood of $\xi=0$ are confined to a $\xi$ interval of size
$O(\frac{1}{\sqrt{K}})$. Beyond this interval
($\frac{1}{\sqrt{K}}\ll \xi<\xi_{max}$) $P(\xi,I_K)$ takes the form
\begin{equation}
P(\xi,I_K) \sim \frac{1}{2} \left (\frac{1}{ 1-\frac{
\xi}{\eta_{-}}}  +\frac{1}{1-\frac{ \xi}{\eta_{+}}}   \right )\ \ ,
\end{equation}
which coincides with the small $\xi$ expression of the limit
distribution (\ref{eq:xismall}) to leading order in $\xi$. To obtain
the dominant behavior of $P(\xi,I_K)$ near $\xi=0$, one should add
$P_{m=0}(\xi,I_K)$ which we compute now.

\noindent \emph{The contribution of the $m=0$ term in (\ref
{eq:pixim})}. Following the same steps as above, the leading
approximation to $P_{m=0}(\xi,I_K )$ reads,
\begin{equation}
\hspace{-25mm} P_{m=0}(\xi,I_K )\sim {1\over g
K}\int_{-\frac{1}{2}}^{\infty}\ \chi_{I_K}\left(2 \mathcal {A}
H(n+\frac{1}{2},0))\right)\ \delta \left(\xi - \frac{n+1}{2 \mathcal
{A} H(n+\frac{1}{2},0)}\right ){\rm d}n.
\end{equation}
This being already a correction term, we are allowed to neglect the
semi-classical correction $\frac{1}{2}$ to $n$ in the argument of
$H(n+\frac{1}{2},0)$. With $ H(n,0)=\left (\frac {\pi n
}{\mathcal{L}}\right )^2$ from (\ref{eq:mzerospect}), we get
\begin{equation}
P_{m=0}(\xi,I_K )\sim {1\over g K}\int_{\frac{\mathcal{L}}{\pi}\sqrt
{\frac{K}{2\mathcal{A}}}}^ {\frac{\mathcal{L}}{\pi}
\sqrt{\frac{K(1+g)}{2\mathcal{A}}}}\  \delta \left (\xi -
\frac{1}{n} \frac{\mathcal{L}^2}{2\mathcal{A}\pi^2}\right ){\rm d}n\
.
\end{equation}
Therefore,
\begin{eqnarray}
  P_{m=0}(\xi,I_K ) \sim\left \{
  \begin{array}{ll}
   \frac{1}{\xi^{ 2}}\ \frac{\eta_{-}}{g K}  &
    {\rm if}\ \ \ \sqrt{\frac{\eta_{-}}{K(1+g)}}
    \le \xi < \sqrt{\frac{\eta_{-}}{K}}\    \\
  \ \ \  0 &  \ \ \ \ \ \ \ \ \ \ \ \ \ \ \ \ {\rm otherwise}
  \end{array}
  \right .  \  ,
 \end{eqnarray}
where $\eta_{-}$ was defined in (\ref {eq:etadef}). Both the shift
of the support away from the origin $\xi=0$ and its size, decrease
as $\frac{1}{\sqrt{K}}$. In its support, $P_{m=0}(\xi,I_K)$ is
bounded between the values $\frac{1}{g}$ and $1+\frac{1}{g}$.
However, although the contribution of this term to the probability
density is $O(1)$, its effect on the probability is
$O(\frac{1}{K})$, or in other words,
$P_{m=0}(\xi,I_K)=O(\frac{1}{K})$ in the weak sense. In this
vicinity, $P_{m=0}(\xi,I_K)$ depends only on a single geometric
parameter, $\eta_-$.  Note that \emph{a priori}, one would have
expected the lower limit of the support of $\bar P(\xi,I_K)$ to be
$O(\frac{1}{K})$, and not $O(\frac{1}{\sqrt{K}})$. This prediction
stems from the definition of the normalized nodal sequence (\ref
{eq:nndom}); for $\mathcal{N}>1$ we have the lower bound
$\xi_{\mathcal{N}}\geq \frac{2}{\mathcal{N}}$, so $P(\xi,I_K)\equiv
0$ for $\xi<\frac{2}{(1+g)K}$, and not for
$\xi<O(\frac{1}{\sqrt{K}})$ as observed.

The oscillatory terms $Q(I_K)$ contribute terms of order
$\frac{1}{\sqrt{K}}$ in the $\xi =0$ vicinity. They originate from a
Gibbs phenomenon and they will be discussed in \ref {appendix2}.
Figures 6. and 7. compare the results of numerical simulations with
the expressions derived above in the vicinity of $\xi= 0$, and the
theory includes also the oscillatory corrections (to be precise,
with the integrated density
$\Pi(\xi,I_K):=\int_0^{\xi}P(\xi',I_K){\rm d}\xi'$).

\begin{figure}[h]
\centering \label{fig:picture6}
\includegraphics[width=4in]{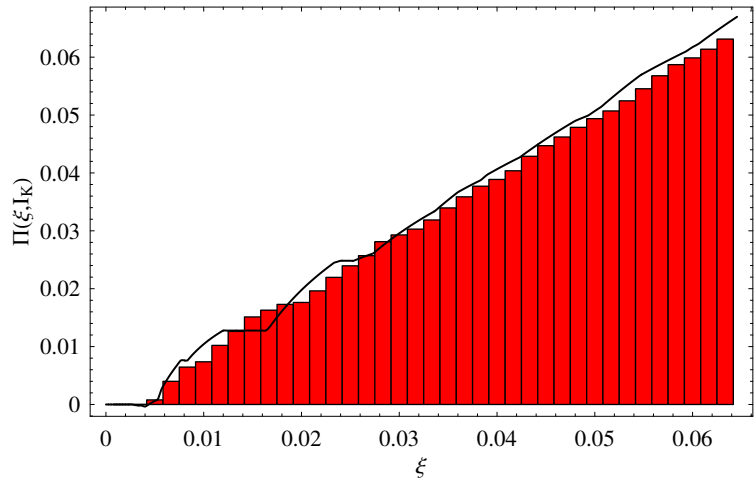}
\caption{\label{full}  Local behavior of $\Pi(\xi,I_K)$ at $\xi=0$
for an ellipsoid of revolution with eccentricity $\varepsilon=0.5$
($K=24000$, $g=1$). }
\end{figure}

\vspace{10mm}

\begin{figure}[h]
\centering \label{fig:picture7}
\includegraphics[width=4in]{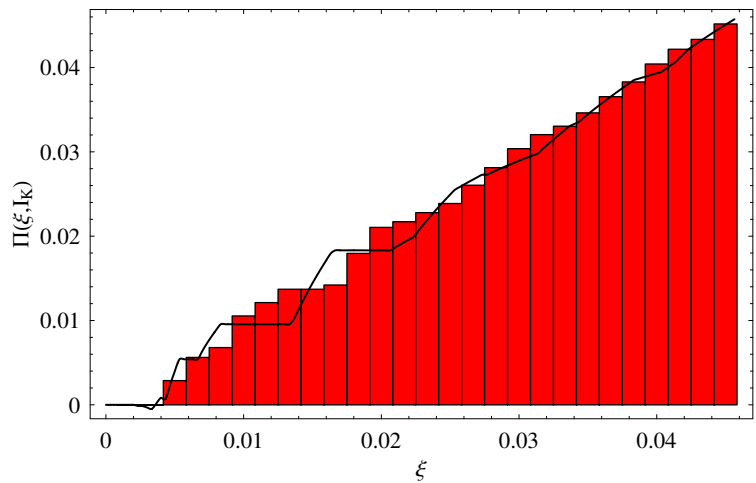}
\caption{\label{full}  Local behavior of $\Pi(\xi,I_K)$ at $\xi=0$
for an ellipsoid of revolution with eccentricity $\varepsilon=0.5$
($K=48000$, $g=0.5$). }
\end{figure}

\subsection{The oscillatory contributions $Q(\xi,I_K)$}
 \label{subsect:oscillatory}

\textit{Weak estimate of $Q(I_K)$}. In this section we  estimate the
oscillatory part $Q(\xi,I_K)$, defined in $(\ref {eq:poisson})$. We
show that it is of smaller order in $\frac{1}{\sqrt{K}}$ than the
leading order correction of the corresponding smooth part. This
justifies the preceding analysis where we considered only the smooth
part, which gives the only contribution of $O(\frac{1}{\sqrt{K}})$.
The numerical results also give evidence of this self-averaging
process in the semi-classical limit.

To be more precise, we shall show that
\begin{equation}
\int_0^1Q(\xi,I_K)\varphi(\xi){\rm d}\xi=o(\frac{1}{K}) \ ,
\end{equation}
for any test function $\varphi$. We shall follow a series of natural
regularizations which are justified in studying this weak limit. We
shall also discuss the local behavior of  $Q(\xi,I_K)$  in the
neighborhood of $\xi=0$, in order to compare with the numerical
results.

We have  $Q(\xi,I_K)=\sum_{(N,M)\neq 0}Q_{N,M}(\xi,I_K)$, where the
Fourier components (\ref{eq:piximne0}) are approximated as
\begin{eqnarray}
\hspace{-25mm} & &Q_{N,M}(\xi,I_K) \sim \nonumber \\
\hspace{-25mm}& &\ \ \ \ \frac{1}{gK}\int_{K/2\mathcal{A}}^{(1+g)K/2\mathcal{A}}
{\rm d}\mathcal{E}\int_{0}^{m_{max}}\delta(\xi-\frac{mn(m)}{\mathcal{A}})
\Big|n(m)-mn'(m)\Big|{\rm e}^{2\pi i\sqrt{\mathcal{E}}(Mm+Nn(m))}{\rm d}m \ .
\end{eqnarray}

We now turn to a smoothing of this distribution by adding a small
imaginary part to the argument of the delta function, say
$\varepsilon>0$. This amounts to replacing the $\delta$ function by
the Lorentzian
\begin{equation}
\delta^{\varepsilon}(\xi)=2\Re\int_{0}^{\infty} {\rm e}^{2\pi i\xi
x}{\rm e}^{-2\pi\varepsilon x} {\rm
d}x=\frac{1}{\pi}\frac{\varepsilon}{\varepsilon^2+\xi^2} \ .
\end{equation}

And so,
\begin{equation}
\hspace{-25mm}Q_{N,M}^{\varepsilon}(\xi,I_K)\sim\frac{\varepsilon}{\pi gK}
\int_{K/2\mathcal{A}}^{(1+g)K/2\mathcal{A}}
\frac{{\rm e}^{2\pi i\sqrt{\mathcal{E}}(Mm+Nn(m))}}
{\varepsilon^2+(\xi-\xi_{n(m),m})^2}\Big|n(m)-mn'(m)\Big|{\rm d}m \ .
\end{equation}

We must notice here a qualitative difference between the
semiclassical theory of the spectral density and the nodal domain
distribution. The index set $I_K=[K,K+\Delta K]$, corresponds to a
spectral interval $[E_K,E_{K+\Delta K}]$. In the later case, as
$K\rightarrow \infty$, the eigenvalues are distributed in an ever
growing interval, while the normalized nodal sequence is distributed
in $(0,1]$, becoming arbitrarily dense. So, in contrast to the
regularization of the spectral density, in general we do not expect
the limits $\varepsilon\rightarrow 0+$ and $K\rightarrow\infty$ to
commute for the nodal domain distribution. We shall first consider
the semi-classical limit.

Once again, by the linear approximation, we have $n(m)=
n_{\pm}+n'_{\pm}(m-m_{\pm})+O(\frac{1}{\sqrt{\mathcal{E}}})$, and
$mn(m)/\mathcal{A}=
\frac{1}{\mathcal{A}}(n_{\pm}+m_{\pm}n'_{\pm})(m-m_{\pm})+
O(\frac{1}{\sqrt{\mathcal{E}}})=:\xi'_{\pm}(m-m_{\pm})+O(\frac{1}{\sqrt{\mathcal{E}}})$.

By taking the whole real line as the $m$-integration range and
shifting $m\mapsto x=m-m_{\sigma}$ ($\sigma=\pm$), we have
$$Q_{N,M}^{\varepsilon}(\xi,I_K)\sim\sum_{\sigma=\pm}
\frac{\varepsilon}{\pi gK}(n_{\sigma}-m_{\sigma}n'_{\sigma})
\int_{K/2\mathcal{A}}^{(1+g)K/2\mathcal{A}}{\rm d}\mathcal{E}
{\rm e}^{2\pi i S\sqrt{\mathcal{E}}}\int_{-\infty}^{\infty}
\frac{{\rm e}^{2\pi i\sqrt{\mathcal{E}}Rx}}
{\varepsilon^2+(\xi-\xi'_{\sigma}x)^2}{\rm d}x=$$
$$=\sum_{\sigma}\frac{1}{gK}
\frac{n_{\sigma}-m_{\sigma}n'_{\sigma}}
{n_{\sigma}+m_{\sigma}n'_{\sigma}}
\int_{K/2\mathcal{A}}^{(1+g)K/2\mathcal{A}}
{\rm e}^{2\pi i S\sqrt{\mathcal{E}}-2\pi\varepsilon\sqrt{\mathcal{E}}
|\frac{R}{\xi_{\sigma}'}|}{\rm d}\mathcal{E}=$$
\begin{equation}
=\sum_{\sigma}\frac{2}{gK}\frac{n_{\sigma}-m_{\sigma}n'_{\sigma}}
{n_{\sigma}+m_{\sigma}n'_{\sigma}}
\int_{\sqrt{K/2\mathcal{A}}}^{\sqrt{(1+g)K/2\mathcal{A}}}
{\rm e}^{2\pi(i S-\varepsilon|\frac{R}{\xi_{\sigma}'}|)k}k{\rm d}k \ ,
\end{equation}
with $\left(\begin{array}{cc}
R \\
S
\end{array} \right)=\left( \begin{array}{cc}
n'_{\sigma} & 1  \\
n_{\sigma} & m_{\sigma}
\end{array} \right)
\left(\begin{array}{cc}
N \\
M
\end{array} \right)$.

Now, instead of a sharp uniform window (taking into account only
those states whose index lie in the interval $I_K$), we consider a
Gaussian regularization,
\begin{equation}
h_K(k)=\frac{1}{\sqrt{\pi}v}\exp\Big(-(k-\mu)^2/v^2\Big)
\end{equation}
with $\mu=\frac{\sqrt{1+g}+1}{2}\sqrt{\frac{K}{2\mathcal{A}}}$ and
$v=(\sqrt{1+g}-1)\sqrt{\frac{K}{2\mathcal{A}}}$, and extend the
integration over the whole real line (of course, this regularization
does not affect the asymptotic behavior we study),
$$Q_{N,M}^{\varepsilon}(\xi,I_K)\sim\sum_{\sigma=\pm}\frac{2}
{\sqrt{\pi}vgK}\frac{n_{\sigma}-m_{\sigma}
n'_{\sigma}}{n_{\sigma}+m_{\sigma}n'_{\sigma}}
\int_{-\infty}^{\infty}
{\rm e}^{2\pi(i S-\varepsilon|\frac{R}{\xi_{\sigma}'}|)k-(k-\mu)^2/v^2}k{\rm d}k=$$
$$=\sum_{\sigma}\frac{2}{\sqrt{\pi}vgK}\frac{n_{\sigma}-m_{\sigma}n'_{\sigma}}{n_{\sigma}
+m_{\sigma}n'_{\sigma}}{\rm e}^{-\mu^2/v^2}\int_{-\infty}^{\infty}
{\rm e}^{-k^2/v^2+(2\pi i S-2\pi\varepsilon|\frac{R}{\xi_{\sigma}'}|+2\mu/v^2)k}k{\rm d}k=$$
$$=\sum_{\sigma}\frac{2v^2}{gK}\frac{n_{\sigma}-m_{\sigma}n'_{\sigma}}
{n_{\sigma}+m_{\sigma}n'_{\sigma}}\Big(\pi i S-\pi
\varepsilon\Big|\frac{R}{\xi_{\sigma}'}\Big|+\frac{\mu}{v^2}\Big)\times$$
\begin{equation}
\times\exp\Big(-\pi^2v^2S^2+\pi^2\varepsilon^2v^2\frac{R^2}
{\xi_{\sigma}^{'2}}-2\pi^2i\varepsilon v^2\Big|\frac{R}{\xi_{\sigma}'}
\Big|S+2\mu \pi iS-2\pi\mu\varepsilon\Big|\frac{R}{\xi_{\sigma}'}\Big|\Big) \ .
\end{equation}

We proceed towards a crude, yet sufficient for our purpose, estimate for $Q(\xi,I_K)$,
$$|Q_{\varepsilon}(\xi,I_K)|\leq\sum_{(N,M)\neq 0}
\sum_{\sigma=\pm}\frac{2v^2}{gK}\Big|\frac{n_{\sigma}-m_{\sigma}n'_{\sigma}}
{n_{\sigma}+m_{\sigma}n'_{\sigma}}\Big|\Big(\pi |S|+\pi
\varepsilon\Big|\frac{R}{\xi_{\sigma}'}\Big|+\frac{\mu}{v^2}\Big)\times$$
\begin{equation}
\times\exp\Big(-\pi^2v^2S^2+\pi^2\varepsilon^2v^2
\frac{R^2}{\xi_{\sigma}^{'2}}-2\pi\mu\varepsilon\Big|\frac{R}{\xi_{\sigma}'}\Big|\Big) \ .
\end{equation}

By taking into account that $\sum_{\gamma\in\mathbb{Z}^2}{\rm
e}^{-t(\alpha\cdot \gamma)^2+\sqrt{t}\beta\cdot
\gamma}\asymp\frac{1}{t}$ and
$\sum_{\gamma\in\mathbb{Z}^2}|\gamma^i|{\rm e}^{-t(\alpha\cdot
\gamma)^2+\sqrt{t}\beta\cdot \gamma}\asymp\frac{1}{t^{3/2}},\textrm{
}t\rightarrow\infty$, we have
\begin{equation}
|Q_{\varepsilon}(\xi,I_K)|\leq O(\frac{1}{K^{3/2}}) \ ,
\end{equation}
uniformly in $\varepsilon$ (before the limit $\varepsilon\rightarrow
0+$ is taken, all three terms in the amplitude of the summed
quantity contribute to this same order; $\frac{\mu}{v^2}\asymp
\frac{1}{\sqrt{K}}$).

\textit{Local estimate of
$Q(\xi,I_K)$ in the neighborhood of $\xi=0$}. From the numerical
investigations, a clear oscillatory behavior of the distribution is
observed in the neighborhood of $\xi=0$, which implies the
importance of the oscillatory part in that region. This is due to a
Gibbs phenomenon, and is not in contradiction to the fact that
the oscillatory part $Q(I_K)$ is of less order in the weak sense
than the smooth part $\bar P(I_K)$ - this is a local contribution.

We find the manifestation in the dominant terms $\sum_N Q_{N,0}$ and
$\sum_{M}Q_{0,M}$, which agrees with the numerical results. The explicit form
of these contributions are presented in \ref {appendix2}.

\section{The geometric information stored in the nodal sequence}
 \label{sect: geometry}
The derivation of the probability density $P(\xi,I_K)$ in the
preceding section is based on functions which are obtained directly
from the dynamical relations embedded in the Hamiltonian and its
dependence on the action variables. These, in turn are computed from
the profile curve $y=f(x)$, which defines $\mathcal{M}$. Here, we
would like to investigate the possibility of inverting this
relationship, and ask what can be said about the surface once the
nodal sequence is known.

A few parameters can be easily extracted from the probability
density $P(\xi,I_K)$. Taking the limit $K\rightarrow \infty$ we get
the limit distribution $P(\xi)$. Its support provides the parameter
$\xi_{max}$. In the vicinity of $\xi=0$, $P(\xi)$ depends
symmetrically on the two parameters $\eta_-$ and $\eta_+$. This
provides a useful relationship, which can be combined with the
information from $P(\xi,I_K)$ near $\xi=0$, for finite $K$, which 
depends on all four (independent) geometric parameters, $\eta_{-},
\eta_{+}, \eta_{0}\ {\rm and}\ \eta_{max}$. These parameters are 
related to the geometry of the surface (\ref {eq:etadef}).

As was commented in the previous section, the relations (\ref
{eq:explicit}) and (\ref{eq:p1inal}), can be solved to obtain the
two branches $m_{-}(\xi)$ and $m_{+}(\xi)$. Consequently, the
function $n(m)$ is determined, since
$n(m_{\pm}(\xi))=\frac{\mathcal{A}\xi} {m_{\pm}(\xi)}$. This will be
the first step towards the inversion of the nodal counting data
which will be explained in the next subsection.

\subsection{Inversion of the Nodal Sequence}
In this section we discuss nodal domain inversion in detail. The
first result is: \emph{with the distribution $\bar P(I_K)$ as given
data, the ``scale-invariant" action $n(m)/\sqrt{\mathcal{A}}$ is
determined uniquely.} For simple surfaces of revolution (separable
systems in general with the Hamiltonian satisfying the homogeneity
condition), $n(m)$ incorporates all the information about the
dynamics, and as we shall see in some cases, the actual geometry.

Our starting point is the semi-classical asymptotic expansion
$(\ref {eq:I1next})$,
\begin{equation}
\bar P(\xi,I_K) \sim P(\xi) + \frac{1}{\sqrt{K}}P_1(\xi) \  ,
\label{eq:inversionseries}
\end{equation}
which, for convenience, we solve for $P_1(\xi)$,
\begin{equation}
P_1(\xi)=\sqrt{K}\Big(\bar P(\xi,I_K)- P(\xi)\Big)+O(\frac{1}{\sqrt{K}}) \  .
\label{eq:identityinv}
\end{equation}
Obviously, $P$ and $P_1$ are determined from the given data,
\begin{equation}
\lim_{K\rightarrow\infty}\bar P(I_K)=P \  ,  \ \  {\rm and}  \ \
\lim_{K\rightarrow\infty}\sqrt{K}\Big(\bar P(I_K)-P\Big)=P_1 \ .
\end{equation}

From relation ($\ref{eq:explicit}$) ,
\begin{equation}
P(\xi)=\xi\frac{{\rm d}}{{\rm d}\xi}\log\frac{m_-(\xi)}{m_+(\xi)}\
\Rightarrow\
\frac{m_+(\xi)}{m_-(\xi)}=\exp\int_{\xi}^{\xi_{max}}\frac{P(\xi')}{\xi'}{\rm
d}\xi' \ , \label{eq:diaforiki1}
\end{equation}
the ratio $m_+(\xi)/m_-(\xi)$ is determined, given $P$. Accompanying
the above, we have the relation (\ref {eq:p1inal}),
\begin{equation}
P_1(\xi)=-\sqrt{\frac{2}{\mathcal{A}}}\frac{\sqrt{1+g}-1}{g}
\sqrt{\xi}\ \frac{{\rm d}}{ {\rm d}\xi}\left( \sqrt{\xi}\ \frac{{\rm
d} } {{\rm d}\xi}\left[m_{+}(\xi)-m_{-}(\xi) \right ]\right )\  \ .
\nonumber
\end{equation}
which  can be inverted to give
\begin{equation}
\hspace{-20mm} u(\xi)\ := \
\frac{1}{\sqrt{\mathcal{A}}}\Big(m_+(\xi)-m_-(\xi)\Big)
=-\frac{1}{\sqrt{2}}\frac{g}{\sqrt{1+g}-1} \int^{\xi}
\frac{1}{\sqrt{\xi'}}\Big(\int^{\xi'}\frac{P_1(\xi'')}{\sqrt{\xi''}}{\rm
d}\xi''\Big){\rm d}\xi' \ . \label{eq:solutions}
\end{equation}
To determine $u(\xi)$ uniquely, two integration constants should be
provided, and they are given  in terms of the initial conditions for
$u(\xi)$ and $\frac{{\rm d} u}{{\rm d}\xi}$ at the point $\xi=0$ or
$\xi=\xi_{max}$. They  are expressed in terms of the geometric
parameters $\{\eta_-,\eta_+,\eta_0,\eta_{max}\}$, which are
extracted from the form of $P(\xi)$ near $\xi=0$ and
$\xi=\xi_{max}$.  Thus the ratio and the difference between
$m_+(\xi)$ are  $m_-(\xi)$ given. Together with
$n_{\pm}(\xi)=\xi\frac{\mathcal{A}}{ m_{\pm}(\xi)}$ they give the
parametric representation of $\Gamma$. Note that we require no
information from the statistical properties of the nodal counts of
the isotropic quantum states - i.e. $P_{m=0}(I_K)$.

The significance of this result becomes apparent in the next
section, where we confine ourselves to mirror-symmetric surfaces.

\subsection {Mirror-symmetric $f(x)$ is uniquely determined by the $n(m)$}
Following the preceding section,  we prove that if the generating
curve $f$ is mirror-symmetric, (i.e. $f(-x)=f(x)$), the action
variable $n(m)$ determines the $f$ uniquely.

Since $f$ is an even function, we may write
\begin{equation}
n(m)=\frac{2}{\pi}\int_{0}^{x_+}\sqrt{f(x)^2-m^2}\frac{\sqrt{1+f'(x)^2}}{f(x)}{\rm d}x \ ,
\end{equation}
where $f(x_+)=m$, or $x_+=f^{-1}(m)$. Changing the variable
$x\mapsto u=f(x)/m_{max}$,
\begin{equation}
n(m)=\frac{2}{\pi}\int_{m/m_{max}}^{1}\frac{\sqrt{m_{max}^2u^2-m^2}}{u}\frac{\sqrt{1+f'\circ
f^{-1}(m_{max}u)^2 }}{|f'\circ f^{-1}(m_{max}u)|}{\rm d}u \ .
\end{equation}

Suppose that $f,g$ are different curves of this class which give the same action variable, i.e.

$$\int_{m/m_{max}(f)}^1\frac{\sqrt{m_{max}(f)^2u^2-m^2}}{u}
\frac{\sqrt{1+f'\circ f^{-1}(m_{max}(f)u)^2 }}{|f'\circ
f^{-1}(m_{max}(f)u)|}{\rm d}u=$$
\begin{equation}
=\int_{m/m_{max}(g)}^1\frac{\sqrt{m_{max}(g)^2u^2-m^2}}{u}\frac{\sqrt{1+g'\circ
g^{-1}(m_{max}(g)u)^2 }}{|g'\circ g^{-1}(m_{max}(g)u)|}{\rm d}u \ ,
\end{equation}
while $f\neq g$.

Firstly, note that $m_{max}(f)=m_{max}(g)$. This is because
$m_{max}$ is the sole real root of $n(m)$, thus, since both sides of
the above equality are proportional to $n(m)$ they must have a
common root, denoted simply be $m_{max}$, since $n(m)$ is analytic
on $(0,m_{max}]$. Thus, since the integration limits are identical,
the integrands must be equal.

It suffices to show that $\frac{\sqrt{1+f'\circ f^{-1}(y)^2
}}{|f'\circ f^{-1}(y)|}=\frac{\sqrt{1+g'\circ g^{-1}(y)^2
}}{|g'\circ g^{-1}(y)|}$, for $y\in[0,m_{max}]$, implies $f=g$. From
the above we have $f'\circ f^{-1}(y)=g'\circ g^{-1}(y)$. The problem
has been reduced to showing that the nonlinear operator $Af=f'\circ
f^{-1}$ acting on our function class, possesses an inverse, i.e.
$Af=Ag\Leftrightarrow f=g$.

Consider the inhomogeneous ``functional" equation $f'\circ
f^{-1}=h$, for some $h$ in some other appropriate function class. 
We shall show that for given $h$, this determines, along with some
initial condition, a unique $f$, formally $f=A^{-1}h$. We have 
$f'(f^{-1}(y))=h(y)$, or $f'(x)=h(f(x))$, since $y=f(x)$. Thus, we 
have reduced this to a first order ordinary differential equation, 
$y'=h(y)$. Accompanied by the initial condition $f(0)=m_{max}$, this 
becomes an initial value problem on $[0,1]$ with a unique 
solution ($h$ is smooth on $[0,1)$). Thus, $A^{-1}$ exists.   

In our problem the initial condition is provided by knowledge of the
root of $n(m)$, so we have reached the conclusion that the two
integrals cannot equal if $f\neq g$.

\section {Summary and conclusions}
The unique inversion of the nodal sequence which was demonstrated
above for symmetric and ``simple" surfaces of revolution paves the
way to a sequence of problems which should now be addressed. Other
families of separable manifolds are known, amongst which the
Liouville surfaces  \cite {Liouville} and the axially symmetric Zoll
surfaces \cite{Zoll} are of prime importance. We believe that the
general approach taken in the present paper could be applied to
handle these case, however, modifications should be applied to take
care of special problems which are intrinsic to these problems, and
this remains for a further study. Another class of integrable
systems consists of independent particle models which are commonly
used in Atomic and Nuclear physics. The study of such systems
extends the research of nodal domains to systems with arbitrary
dimensions.

The next systems in complexity are systems which are classically
integrable but are not separable quantum mechanically. The simplest
examples consist of e.g. the Dirichlet Laplacians in the equilateral
or the isosceles right triangles. Even though the spectrum can be
expressed precisely in terms of the ``quantum numbers", counting of
nodal domains is difficult, and the study of the nodal sequences in
such cases might call for other approaches then the one pursued
here.

Do nodal sequences in other systems store geometric information? Is
there a way to extract this information  to determine the geometry?
These are yet open problems, and the only hint for an affirmative
answer comes from preliminary numerical simulations which indicate
that  ``nodal" trace formulae exist  for ``quantum graphs" \cite
{Rami} and for ``chaotic billiards" \cite {Amit}. No rigorous
treatment exists so far.

\section {Acknowledgments}
The authors would like to thank Prof. Jon Keating and Dr. Sven
Gnutzmann for lengthy and enlightening discussions. This work was
supported by the Minerva Center for non-linear Physics and the
Einstein (Minerva) Center at the Weizmann Institute, by the ISF and
by grants from the  GIF (grant I-808-228.14/2003), and EPSRC (grant
GR/T06872/01).

\appendix
\section {The regularization of the action integral}
\label{appendix1}
The action variable (\ref {eq:action}) is defined
in terms of an integral, whose form is not convenient for further
computations, such as e.g., the evaluation of its higher
derivatives. This can be done by regularizing the integral in a way
which will be explained here. Rather than introducing fractional
derivatives as was done in e.g., \cite{Gurarie}, we compute the
integrals explicitly. It is convenient to introduce the notation
$q(x)= f(x)^2$ and $\mu=m^2$. We start with
\begin{equation}
\hspace{-1cm}
 n(\mu) = \frac{1}{\pi} \int_{x_{-}}^{x_{+}}\sqrt{ q(x)-\mu}
\frac{\sqrt{4 q(x)+q'(x)^2}}
{2q(x)}\ {\rm d}x \ , \ {\rm where}\ \ q(x_{\pm})=\mu \ .
 \label{eq:action1}
\end{equation}
The function $q(x)$ is analytic in $I$ and has a single maximum at
$x_{max}$.  We separate the integration interval in
(\ref{eq:action1}) to two consecutive intervals $[x_{-},x_{max}]$
and $[x_{max},x_{+}]$ and write accordingly
  \begin{equation}
 n(\mu) =\  n_{-}(\mu) \ +\ n_{+}(\mu) \ .
\label{eq:action2}
\end{equation}
In each of the intervals $[-1,x_{max})$ and $(x_{max},1]$, $q(x)$ is
a monotonic function. Therefore it  can be inverted in each of the
intervals in terms of the corresponding functions $x_{\pm}(q)$,
which are analytic in the interval $[0,q_{max})$. Thus,
\begin{equation}
n_{\pm}(\mu) =\frac{1}{\pi}\int_{\mu}^{q_{max}} \sqrt{
q-\mu } w_{\pm}(q) {\rm d}q \label{eq:fin1}
\end{equation}
 with
\begin{equation}
  w_{\pm}(q) = \frac{1} {2q }\sqrt{ 4 q \left
(\frac {{\rm d}x_{\pm}(q)}{{\rm d}q}\right )^2+1} .
 \label{eq:fin2}
\end{equation}
The expressions in the square brackets above are  analytic in the
domain of $q$ where $x_{\pm}(q)$ are analytic, that is, in
$[0,q_{max})$. They can be Taylor expanded with a convergence radius
$q_{max}$ so that
\begin{equation}
w_{\pm}(q) = \frac{1}{2q} + \sum_{r=0}^{\infty} \tau^{\pm}_r q^{r}
\ . \label{eq:wexpand}
\end{equation}
Substituting this expressions in (\ref{eq:fin1}), performing the
integrals and defining $ \tau_r =\tau^{+}_r+\tau^{-}_r$, we finally
get
\begin{equation}
\hspace{-1cm}
 n(\mu)\ =\frac{2}{\pi} \left [\left( q_{max}-\mu
\right )^{\frac{1}{2}} - \mu^ {\frac{1}{2}}  \arccos \left (
\frac{\mu}{q_{max}}\right )^ {\frac{1}{2}}\right ]\ +\
\frac{1}{\pi}\sum_{r=0}^{\infty} \tau_r\ I_r(\mu) \ ,
\end{equation}
where,
\begin{eqnarray}
\hspace{-1.5cm} I_r(\mu ) = \int_{\mu}^{q_{max}}
q^r\sqrt{ q -\mu }{\rm d}q  = (q_{max}-\mu
)^{\frac{3}{2}}\sum_{k=0}^r  {r\choose k} \frac
{(q_{max}-\mu)^{r-k}\ \mu^k}{\ (r-k)+ \frac{3}{2}\ }\ .
\end{eqnarray}
Using (\ref {eq:nof0}) we find
\begin{equation}
 n(0)\ =\frac{\mathcal{L}}{\pi}=\frac{1}{\pi}
 \sqrt{q_{max}}\left [2 + \sum_{r=0}^{\infty}  \frac { \tau_r\ q^{\
r+1}_{max}}{r +\frac{3}{2}} \right] .
\end{equation}
Thus in the vicinity of $\mu=0$ we get
\begin{equation}
n(\mu)\ \sim \frac{\mathcal{L}}{\pi} -\sqrt{\mu}\ .
 \label{eq:smallmu}
\end{equation}
The behavior of $n(\mu)$ near the other extreme end of the interval
- $\mu = q_{max}$ - cannot be deduced in the same way, because
$x(q)$ is not defined at this point. However, staring directly from
(\ref{eq:action1}) we can obtain the behavior of $n(\mu)$ in this
domain. For this purpose we write
\begin{equation}
q(x) \sim q_{max} - \frac{1}{2} \omega\ (x-x_{max})^2 \ ,\
\omega =|q''(x_{max})| \ \ , \ \ x \rightarrow x_{max} \ .
\end{equation}
To leading order in $(q_{max}-\mu)$, (\ref{eq:action1}) reduces to
\begin{equation}
 \hspace{-1.5cm}
n(\mu) \sim \frac{2}{\pi \sqrt{2 \omega q_{max}}}
\int_{\mu}^{q_{max}}\sqrt{\frac{q-\mu}{q_{max}-q}}{\rm d}q \ = \
\frac{q_{max}-\mu}{\sqrt{2 \omega q_{max}}}\ \sim
{\sqrt\frac{2}{\omega}}(f_{max}-|m|).
 \label{eq:lrgmu}
\end{equation}

Finally, we return to the example of the sphere. With $f(x)^2
=1-x^2$ the integrals can be performed exactly,
\begin{equation}
n(m) = 1-|m| \ .
 \label{eq:nsphere}
\end{equation}
This is consistent with the local expressions presented in (\ref {eq:smallmu},\ref{eq:lrgmu}).

\section {The oscillatory part near the origin}
\label{appendix2}
 We recover the oscillatory behavior of the
distribution near $\xi=0$ from the leading order sums
\begin{equation}
Q(\xi,I_K)\sim\sum_{N\in\mathbb{Z}_*}Q_{N,0}(\xi,I_K)+\sum_{M\in
\mathbb{Z}_*}Q_{0,M}(\xi,I_K) \ .
\end{equation}
We begin with $(\ref{eq:piximne0})$. As $\xi\rightarrow 0+$, the
contribution in the above integral will come from the neighborhouds
of $m=0$ and $m=m_{max}$ (the delta function of the integrand is
supported on $\{m+\delta m\}_{\pm}$). These will be treated
separately, denoted by $Q^{\pm}_{N,M}(\xi,I_K)$ respectively, so
that
\begin{equation}
Q_{N,M}(\xi,I_K)= Q^-_{N,M}(\xi,I_K)+Q^+_{N,M}(\xi,I_K) \  .
\end{equation}
In what follows, we define $x:=\xi/\eta_-$ when referring to the
$Q^-$ terms, and $x:=\xi/\eta_+$ for the $Q^+$ terms.

Given that $m_-\sim \frac{\mathcal{L}}{2\pi}x$ and $\delta m_-\sim
-\frac{1}{2}\eta x$, by changing variables to the appropriate
dimensionless wavenumber $k:=\mathcal{L}\sqrt{\mathcal{E}}$, we have
\begin{eqnarray}
\hspace{-25mm} & &Q_{0,M}^-(\xi,I_K) \sim \nonumber \\
\hspace{-25mm}& &\ \ \ \ \frac{1}{\pi^2\eta _-gK}\frac{{\rm e}^{-\pi
i
Mx/2}}{1-x}\int_{\pi\sqrt{\eta_-K}}^{\pi\sqrt{\eta_-(1+g)K}}\Theta\Big(k-\frac{\pi}{x}\Big){\rm
e}^{i M x k}\Big(k-\frac{\pi}{2}\Big){\rm d}k \ .
\end{eqnarray}
Similarly, we carry out the calculation of the integral
$Q_{0,M}^+(\xi,I_K)$. Here, the condition that $(m+\delta m)_+$ lies
in the integration range reads $\mathcal{A}\xi>\Big(mn(m)+\eta
m\Big)_{m_{max}}=m_{max}\eta$,
\begin{eqnarray}
\hspace{-25mm} & &Q_{0,M}^+(\xi,I_K) \sim \nonumber \\
\hspace{-25mm}& &\ \ \ \ \frac{1}{\eta  _0gK}\frac{{\rm e}^{\pi i
M\sqrt{\omega/2}(1+x/2)}}{1-x}\int_{\sqrt{\eta_0K}}^{\sqrt{\eta_0(1+g)K}}\Theta\Big(k-\frac{1}{x}\Big){\rm
e}^{2\pi i M (1-x/2) k}\Big(k-\frac{\sqrt{\omega/2}}{2}\Big){\rm d}k
\ .
\end{eqnarray}
where the appropriate wavenumber is $k:=m_{max}\sqrt{\mathcal{E}}$.

Following the above calculations,
\begin{eqnarray}
\hspace{-25mm} & &Q_{N,0}^-(\xi,I_K) \sim \nonumber \\
\hspace{-25mm}& &\ \ \ \ \frac{1}{\pi^2\eta _-gK}\frac{{\rm e}^{\pi
i
Nx/2}}{1-x}\int_{\pi\sqrt{\eta_-K}}^{\pi\sqrt{\eta_-(1+g)K}}\Theta\Big(k-\frac{\pi}{x}\Big){\rm
e}^{i N(2-x)k}\Big(k-\frac{\pi}{2}\Big){\rm d}k \ ,
\end{eqnarray}
and
\begin{eqnarray}
\hspace{-25mm} & &Q_{N,0}^+(\xi,I_K) \sim \nonumber \\
\hspace{-25mm}& &\ \ \ \ \frac{1}{\eta _0gK}\frac{{\rm e}^{-\pi i
N(1+x/2)}}{1-x}\int_{\sqrt{\eta_0K}}^{\sqrt{\eta_0(1+g)K}}\Theta\Big(k-\frac{1}{x}\Big){\rm
e}^{\pi i N
\sqrt{\omega/2}xk}\Big(k-\frac{\sqrt{2/\omega}}{2}\Big){\rm d}k \ .
\end{eqnarray}

By performing the integrations, we have
\begin{eqnarray}
\hspace{-25mm}\sum_{M\in\mathbb{Z}_*}Q_{0,M}(\xi,I_K)  \sim
\frac{1}{\pi\eta_-gK}\frac{1}{(1-x)x^2}\nonumber \left \{
\begin{array}{ll}
0 &  {\rm for}\ \sqrt{\frac{\eta_{-}}{(1+g)K}}\geq\xi\\
f_1^-(\xi,I_K)& {\rm for} \ \sqrt{\frac{1}{(1+g)\eta_-K}}<\xi\leq\sqrt{\frac{1}{\eta_-K}}\\
f_2^-(\xi,I_K)& {\rm for}\ \sqrt{\frac{1}{\eta_{-}K}}<\xi \nonumber
\end{array}\right .\\
\hspace{-25mm}+\frac{1}{\pi\eta_0gK}\frac{1}{(1-x)(2-x)^2} \left
\{\begin{array}{ll}
0 &  {\rm for} \ \sqrt{\frac{\omega}{2(1+g)\eta_0K}} \geq  \xi\\
f^+_1(\xi,I_K) & {\rm for} \ \sqrt{\frac{\omega}{2\eta_0(1+g)K}} < \xi\leq\sqrt{\frac{\omega}{2\eta_0K}}\\
f^+_2(\xi,I_K) & {\rm for} \ \sqrt{\frac{\omega}{2\eta_0 K}} < \xi \
, \nonumber
\end{array}\right.
\end{eqnarray}
where
$$f^-_1(\xi,I_K):=\alpha(1-\frac{x}{2})-\alpha(x\sqrt{\eta_-K}-\frac{x}{2})+x\sqrt{\eta_-K}\beta(x\sqrt{\eta_-K}-\frac{x}{2})-\beta(1-\frac{x}{2}) \ ,$$
$$f^-_2(\xi,I_K):=\alpha(x\sqrt{(1+g)\eta_-K}-\frac{x}{2})-\alpha(x\sqrt{\eta_-K}-\frac{x}{2})+$$
\begin{equation}
+x\sqrt{\eta_-K}\beta(x\sqrt{\eta_-K}-\frac{x}{2})-x\sqrt{(1+g)\eta_-K}\beta(x\sqrt{(1+g)\eta_-K}-\frac{x}{2})
\ ,
\end{equation}
and
$$f^+_1(\xi,I_K):=\alpha\Big(\frac{2}{x}-1+(1+\frac{x}{2})\sqrt{\frac{\omega}{2}}\Big)-\alpha\Big((2-x)\sqrt{\eta_0K}+(1+\frac{x}{2})\sqrt{\frac{\omega}{2}}\Big)+$$
$$+2\sqrt{\eta_0K}\beta\Big((2-x)\sqrt{\eta_0K}+(1+\frac{x}{2})\sqrt{\frac{\omega}{2}}\Big)-\frac{2}{x}\beta\Big(\frac{2}{x}-1+(1+\frac{x}{2})\sqrt{\frac{\omega}{2}}\Big) \ ,$$

$$f^+_2(\xi,I_K):=\alpha\Big((2-x)\sqrt{(1+g)\eta_0K}+(1+\frac{x}{2})\sqrt{\frac{\omega}{2}}\Big)-\alpha\Big((2-x)\sqrt{\eta_0K}+(1+\frac{x}{2})\sqrt{\frac{\omega}{2}}\Big)+$$
$$+2\sqrt{\eta_0K}\beta\Big((2-x)\sqrt{\eta_0K}+(1+\frac{x}{2})\sqrt{\frac{\omega}{2}}\Big)-$$
\begin{equation}
-2\sqrt{(1+g)\eta_0K}\beta\Big((2-x)\sqrt{(1+g)\eta_0K}+(1+\frac{x}{2})\sqrt{\frac{\omega}{2}}\Big)
\ .
\end{equation}

Similarly,
\begin{eqnarray}
\hspace{-25mm}\sum_{N\in\mathbb{Z}_*}Q_{N,0}(\xi,I_K) \sim
\frac{1}{\pi\eta_-gK}\frac{1}{(1-x)(2-x)^2}\nonumber \left
\{\begin{array}{ll}
0 &  {\rm for}\ \sqrt{\frac{\eta_{-}}{(1+g)K}}\geq\xi\\
g^-_1(\xi,I_K)& {\rm for} \ \sqrt{\frac{1}{(1+g)\eta_-K}}<\xi\leq\sqrt{\frac{1}{\eta_-K}}\\
g^-_2(\xi,I_K)& {\rm for}\ \sqrt{\frac{1}{\eta_{-}K}}<\xi \nonumber
\end{array}\right .\\
\hspace{-25mm}+\frac{\omega}{2\pi\eta_0gK}\frac{1}{(1-x)x^2} \left
\{
\begin{array}{ll}
0 &  {\rm for} \ \sqrt{\frac{\omega}{2(1+g)\eta_0K}} \geq  \xi\\
g^+_1(\xi,I_K) & {\rm for} \ \sqrt{\frac{\omega}{2\eta_0(1+g)K}}
 <\xi\leq\sqrt{\frac{\omega}{2\eta_0K}}\\
g^+_2(\xi,I_K) & {\rm for} \ \sqrt{\frac{\omega}{2\eta_0 K}} < \xi \
, \nonumber
\end{array}\right .\\
\end{eqnarray}
where
$$g^-_1(\xi,I_K):=\alpha(\frac{2}{x}-1+\frac{x}{2})-\alpha\Big((2-x)\sqrt{\eta_-K}+\frac{x}{2}\Big)+$$
$$+2\sqrt{\eta_-K}\beta\Big((2-x)\sqrt{\eta_-K}+
\frac{x}{2}\Big)-\frac{2}{x}\beta(\frac{2}{x}-1+\frac{x}{2}) \ ,$$
$$g^-_2(\xi,I_K):=\alpha\Big(\frac{x}{2}+(2-x)\sqrt{(1+g)\eta_-K}\Big)-\alpha\Big(\frac{x}{2}+(2-x)\sqrt{\eta_-K}\Big)+$$
\begin{equation}
\hspace{-25mm}+2\sqrt{\eta_-K}\beta\Big((2-x)\sqrt{\eta_-K}+\frac{x}{2}\Big)-2\sqrt{(1+g)\eta_-K}\beta\Big((2-x)\sqrt{(1+g)\eta_-K}+\frac{x}{2}\Big)\
,
\end{equation}
and
$$\hspace{-25mm}g^+_1(\xi,I_K):=\alpha(\sqrt{\frac{2}{\omega}}-1-\frac{x}{2})-\alpha\Big(x\sqrt{\frac{2\eta_0}{\omega} K}-(1+\frac{x}{2})\Big)+$$
$$+x\sqrt{\frac{2\eta_0}{\omega}K}\beta\Big(x\sqrt{\frac{2\eta_0}{\omega}K}-(1+\frac{x}{2})\Big)-\sqrt{\frac{2}{\omega}}
\beta\Big(\sqrt{\frac{2}{\omega}}-(1+\frac{x}{2})\Big) \ ,$$
$$\hspace{-25mm}g^+_2(\xi,I_K):=\alpha\Big(x\sqrt{(1+g)\frac{2\eta_0}{\omega}K}-(1+\frac{x}{2})\Big)-\alpha\Big(x\sqrt{\frac{2\eta_0}{\omega}K}-(1+\frac{x}{2})\Big)+$$
\begin{equation}
\hspace{-25mm}+\sqrt{\frac{2\eta_0}{\omega}K}x\beta\Big(x\sqrt{\frac{2\eta_0}{\omega}K}-(1+\frac{x}{2})\Big)-\sqrt{(1+g)\frac{2\eta_0}{\omega}K}x\beta\Big(x\sqrt{(1+g)\frac{2\eta_0}{\omega}K}-(1+\frac{x}{2})\Big)
\ .
\end{equation}

Above, we have made use of the periodic functions $\alpha$ and
$\beta$, which we define via their Fourier series,
\begin{equation}
\alpha(t):=\frac{1}{\pi}\sum_{n\in\mathbb{Z}_*}\frac{{\rm e}^{\pi i
nt}}{n^2}=\frac{2}{\pi}\sum_{n\in\mathbb{N}}\frac{\cos \pi n t}{n^2}
\ ,
\end{equation}
and
\begin{equation}
\beta(t):=i\sum_{n\in\mathbb{Z}_*}\frac{{\rm e}^{\pi i
nt}}{n}=-2\sum_{n\in\mathbb{N}}\frac{\sin \pi n t}{n} \ .
\end{equation}

\end{document}